\documentclass[floatfix,lengthcheck,showpacs,amssymb,amsmath,amsfonts,twocolumn]{aastex63}
\usepackage{CJK}
\usepackage{lineno}
\usepackage[utf8]{inputenc}
\usepackage{color}
\usepackage{graphicx}
\usepackage{float}
\usepackage{subfigure}
\usepackage{physics}
\usepackage{mathtools}
\usepackage{amsmath}
\usepackage{afterpage}
\usepackage{acronym}
\usepackage{multirow}
\usepackage{tabularx}
\usepackage{booktabs}
\usepackage{tikz}
\usetikzlibrary{arrows,shapes,trees,decorations.pathreplacing}
\usepackage{import}
\usepackage{svn-multi}
\usepackage{lineno}
\usepackage{hyperref}
\usepackage{gensymb}
\usepackage{bm}
\hypersetup{
    colorlinks=true,
    linkcolor=blue,
    filecolor=magenta,      
    urlcolor=cyan,
}

\newcommand{\AEI}{\affiliation{Max-Planck-Institut f{\"u}r Gravitationsphysik (Albert-Einstein-Institut), D-30167 Hannover, Germany}}
\newcommand{\UniHannover}{\affiliation{Leibniz Universit{\"a}t Hannover, D-30167 Hannover, Germany}}

\begin{document}
\begin{CJK*}{UTF8}{gbsn}
\title[]{Probing Cosmology with Baryon Acoustic Oscillations using Gravitational Waves}
\correspondingauthor{Sumit Kumar}
\email{sumit.kumar@aei.mpg.de}
\author[0000-0002-6404-0517]{Sumit Kumar}
\AEI{}
\UniHannover{}


\begin{abstract}
The third-generation (3G) gravitational wave (GW) detectors such as the Einstein telescope (ET) or Cosmic Explorer (CE) are expected to play an important role in cosmology. With the help of 3G detectors, we will be able to probe large-scale structure (LSS) features such as baryon acoustic oscillations (BAO), galaxy bias, etc. We explore the possibility to do precision cosmology, with the 3G GW detectors by measuring the angular BAO scale using localization volumes of compact binary merger events. Through simulations, we show that with a 3G detector network, by probing the angular BAO scale using purely GW observations, we can constrain the Hubble constant for the standard model of cosmology ($\Lambda$CDM) with $90\%$ credible regions as $H_0 = 59.4^{+ 33.9}_{-17.7} ~\mathrm{km}~\mathrm{s}^{-1}~\mathrm{Mpc}^{-1}$. When combined with BAO measurements from galaxy surveys, we show that it can be used to constrain various models of cosmology such as parametrized models for dark energy equations of state. 
We also show how cosmological constraints using BAO measurements from GW observations in the 3G era will complement the same from spectroscopic surveys.
\end{abstract}

\section{Introduction}
In the last few years, the detection of gravitational waves (GW) from the merger of compact objects has become a routine \citep{LIGOScientific:2016vbw, PhysRevLett.119.161101} and results in detailed catalogs of gravitational wave mergers \citep{ LIGOScientific:2021djp, nitz20214ogc}. The growth of the catalogs enabled us to probe various aspects of science, to list a few: i) inferring the population properties, such as the mass, spin and redshift distribution of the compact binaries \citep{LIGOScientific:2021psn}, ii) testing the validity of general relativity \citep{LIGOScientific:2021sio}, iii) constraining the equation of state and radii of neutron stars \citep{LIGOScientific:2018cki,Capano:2019eae}, iv) constraining the cosmic expansion history and inferring the value of the Hubble parameter \citep{LIGOScientific:2021aug}, etc.

The idea of probing cosmology with GWs is not only exciting but also timely as there exists a tension between the value of the Hubble constant $H_0$ measured from low redshift data such as supernovae (SNe) \citep{Riess:2019cxk} and the data from surveys from the high redshift such as the cosmic microwave background (CMB) \citep{Planck:2018vyg}. For example, the value of $H_0$ as obtained from Planck 2018 results indicate $H_0 = 67.04\pm 0.5~\mathrm{km}~\mathrm{s}^{-1}~\mathrm{Mpc}^{-1}$ \citep{Planck:2018vyg}, while the inferred value from the low redshift probes such as SNIa yields $H_0 = 74.03 \pm 1.42~\mathrm{km}~\mathrm{s}^{-1}~\mathrm{Mpc}^{-1}$  \citep{Riess:2019cxk}. A recent measurement of the local value of $H_0$ from the Hubble Space Telescope (HST) and SHOES team provides a constraint on the $H_0$ with $\sim 1~\mathrm{km}~\mathrm{s}^{-1}~\mathrm{Mpc}^{-1}$ uncertainty as $H_0=73.30\pm 1.04~\mathrm{km}~\mathrm{s}^{-1}~\mathrm{Mpc}^{-1}$, which implies a $5\sigma$ difference with the value predicted by \emph{Planck} 2018 measurements \citep{Riess:2021jrx}. All these measurements assume the standard model of cosmology known as the $\Lambda$CDM model.

The very first detection of GWs from the merger of a binary neutron star (BNS) \citep{PhysRevLett.119.161101} was accompanied by observations from various electromagnetic (EM) telescopes \citep{LIGOScientific:2017ync, LIGOScientific:2017zic}, which made it possible to put the very first constraints on the value of the Hubble constant from GW observations \citep{LIGOScientific:2017adf}. Since then, binary black hole (BBH) merger events have also been used to put constraints on the Hubble parameter, by cross-correlating their localization volumes with the galaxy catalogs \citep{LIGOScientific:2019zcs}.  The degeneracy between the mass and redshift of observed BBH mergers was also explored along with the population models to put constraints on the value of $H_0$ \citep{Mastrogiovanni:2021wsd}. The recent estimates of the $H_0$ from recent GWTC-3 catalog with $68\%$ CL indicates $H_0=68^{+8}_{-6}~\mathrm{km}~\mathrm{s}^{-1}~\mathrm{Mpc}^{-1}$ \citep{LIGOScientific:2021aug}. Though the uncertainties on the value of $H_0$ measured from the GW observations is not at the level of resolving Hubble tension right now, we expect that in the near future, with more GW observations and improvements in the detector sensitivity, we can resolve this tension \citep{Chen:2020zoq}.

The current generation of detectors such as LIGO-Hanford, LIGO-Livingston \citep{LIGOScientific:2014pky}, Virgo \citep{VIRGO:2014yos}, and KAGRA \citep{KAGRA:2020tym} are set to undergo upgrades in various stages in upcoming years \citep{KAGRA:2013rdx}, and new detectors such as LIGO-India \citep{Saleem:2021iwi} are expected to join the global detector network at some point in the future. Thanks to these network improvements, the source localization is expected to improve considerably \citep{Fairhurst:2012tf}. Furthermore, the proposed third-generation (3G) ground-based detectors such as Cosmic Explorer (CE) \citep{Reitze:2019iox, 2021arXiv210909882E} and the Einstein telescope (ET) \citep{Sathyaprakash:2012jk, Punturo_2010} are expected to be operational sometime during the next decade. These detectors will be able to probe the Universe up to very high redshifts ($z\sim10$) and will be able to detect thousands of GW merger events per year \citep{Mills:2017urp, Borhanian:2022czq}. Many of these GW mergers (at lower redshifts) are expected to be localized within a square degree so that the spatial distribution of the localization volumes of well-localized mergers can be used to probe the large-scale structure (LSS) of the Universe, \emph{e.g.} by measuring the galaxy bias \citep{Vijaykumar:2020pzn}, or by detecting the baryon acoustic oscillations (BAO) peak \citep{Kumar:2021aog}, solely from the GW observations. The evolution of the galaxy bias as a function of redshift can be used to do precision cosmology with GW merger events \citep{Mukherjee:2020hyn}.

In this work, we explore another aspect of probing cosmology with the 3G GW detectors through the LSS. We show that by detecting the angular BAO peak using localization volumes of mergers with the 3G GW detector network, we can put independent constraints on the value of $H_0$. Moreover, by combining these results with the BAO measurements from the galaxy surveys, we should be able to put stringent constraints on various cosmological parameters for the standard model of cosmology, as well as on other phenomenological models for the dark energy parametrization. The BAO measurements from the spectroscopic surveys such as SDSS do not constrain the Hubble parameter $H_0$ on its own. In order to put constraints on $H_0$, the BAO measurements need to be combined with other observations such as SNIa, CMB data, etc \citep{eBOSS:2020yzd}. On the other hand, the BAO measurements, that will be obtained from the localization volumes of mergers of compact binary coalescence (CBC) sources with the 3G detector network, will have capabilities to constrain the Hubble parameter $H_0$ on its own. We show that by combining BAO observations from spectroscopic surveys and 3G GW observations, we will be able put combined constraints on the cosmological model as both data sets are complementary to each other.

The structure of this paper is as follows. In section \ref{sec:gwcosmology}, we outline the existing methods to probe cosmology using the current and next generation of GW detectors. We also lay down the methodology to use the BAO measurements with GW merger events, to constrain cosmological models. In section \ref{sec:simulations}, we use simulated data to apply these methods to constraint dark energy (DE) models. We use three parametrized DE models along with the standard $\Lambda$CDM model. In section \ref{sec:summary}, we summarize the results.

\section{Cosmology with gravitational waves}
\label{sec:gwcosmology}
The data from various cosmological surveys indicate that at present, the major constituents of the Universe are dark energy, dark matter, and baryonic matter \citep{Planck:2018vyg}. One of the simplest models which describes the Universe is the so-called $\Lambda$CDM model, which interprets the dark energy component of the Universe in terms of the presence of a cosmological constant $\Lambda$ term in the Einstein equations, along with cold dark matter (CDM), and the baryonic matter which represents all visible matter in the Universe \citep{riess1998observational, Planck:2018vyg, weinberg2013observational}. 

The data from GW detectors consists of a time series $s(t)$ which contains noise $n(t)$ and might contain a GW signal $h(t)$. The GW signal from the merger of two compact objects is modelled as a function of the intrinsic parameters such as individual masses and spins, as well as extrinsic parameters such as the luminosity distance ($D_L$), the inclination angle of the binary with respect to the line of sight, the sky localization (right ascension and declination angles), etc. The localization volumes estimated for a GW event provide a posterior distribution on sky location (RA, dec) and $D_L$. If, somehow, we can estimate the redshift ($z$) of the GW event independently \citep{Holz:2005df, Dalal:2006qt, Nissanke:2013fka}, then using the $D_L - z$ relation from the so-called Hubble equation, we can put constraints on the parameters of the cosmological model, such as the Hubble parameter ($H_0$), the density parameter corresponding to the matter component ($\Omega_{m0}$), etc. The first detection of gravitational waves from the merger of binary neutron stars, known as GW170817 \citep{PhysRevLett.119.161101} provided one such opportunity. The electromagnetic (EM) afterglow of GW170817 was measured by various telescopes across the globe. It provided the constraints on $H_0$ using a GW event for the first time \citep{LIGOScientific:2017adf}. Since then, various schemes have been used to probe cosmology by GW observations. For binary black holes (BBH) merger events, the localization posteriors can be cross-correlated with the galaxy catalogs to put constraints on $H_0$ \citep{LIGOScientific:2019zcs}. Other methods exploit the degeneracy between the inferred component masses from GW events and their redshift by putting combined constraints on $H_0$ and on the population parameters \citep{Mastrogiovanni:2021wsd}.

\subsection{Baryon Acoustic Oscillations}
BAO are imprints on the distribution of matter from the very early Universe. In the standard model of cosmology, the evolution of the Universe is described through three major phases where the dominant component is radiation, matter, and dark energy respectively. In the very early time, the Universe is assumed to have gone through a period of rapid accelerated expansion, known as inflation, resulting in an extremely homogeneous Universe \citep{Guth:1980zm, Linde:1981mu, Baumann:2009ds}. After this period, the Universe enters what is known as the radiation-dominated era, when the temperature of the Universe was very high, so that the protons and electrons could not form a stable hydrogen atom. The Universe was dominated by dark matter, and a hot plasma soup of electrons, protons, and photons. The small perturbations of Gaussian nature in the very early Universe acted as seeds for inhomogeneities and those perturbations grew with time. The competing forces between gravity and electromagnetic radiation pressure in the fluid generated the perturbations which act as sound waves in the hot plasma. About 380,000 years after the big bang, when the temperature of the Universe dropped to a level such that the electrons and protons could combine to form hydrogen atoms, the photons are set free, known as the cosmic microwave background (CMB), and the sound waves were frozen \citep{Hu:2001bc, Planck:2018vyg}. These features have been preserved in the distribution of matter as the Universe evolved. These imprints are called Baryon acoustic oscillations (BAO) \citep{Bassett:2009mm, Weinberg:2013agg} and can be seen in the two-point correlation function (2PCF) estimated from the distribution of galaxies \citep{1980lssu.book.....P, Landy:1993yu,SDSS:2005xqv}. The comoving sound horizon or BAO scale: $r_s$ corresponds to the distance sound waves traversed before they become frozen. The first confident detection of this BAO feature with $3.4\sigma$ certainty was reported by the Sloan Digital Sky Survey (SDSS) data release 3 \citep{SDSS:2005xqv} by measuring 2PCF of the luminous red galaxies. The BAO scale $r_s$ can be used to probe the cosmology as it provides a standard ruler.\\
\\
\subsection{Cosmology using the Large-Scale Structures of the Universe}
The large-scale structures (LSS) ($>\mathcal{O}(10~\mathrm{Mpc})$) of the Universe can be studied by probing the distribution of matter, such as in the galaxy surveys, using the 2PCF $\xi(r)$, which is related to the excess probability $\delta P$ with respect to the expected random distribution, of finding a pair of galaxies separated by a distance $r$,
\begin{equation}
    \delta P(r) = n[1+\xi(r)]dV, \label{equqtion:excess_probability}
\end{equation}
\noindent
where $n$ is the average number of galaxies per unit volume and $dV$ is the infinitesimal volume or volume element around a galaxy. The 2PCF $\xi(r)$ can be estimated from the matter overdensity field $\delta(\textbf{x}) \coloneqq \rho(\textbf{x})/\overline{\rho} - 1$, where $\rho(\textbf{x})$ is the local matter density at position $\textbf{x}$ and $\overline{\rho}$ is the average matter density of the Universe, as
\begin{equation}
    \xi(r) = \left\langle \, \delta(\textbf{x}) \delta(\textbf{y}) \, \right\rangle, \label{equation:2pcf}
\end{equation}
\noindent
where the operation $\left\langle \, \cdot \, \right\rangle$ represents the ensemble average over a large volume compared to the scales we are probing. An important assumption here is the statistical homogeneity and isotropy of the Universe. Due to these assumptions, the correlation function $\xi$ depends only on the magnitude of the separation between points $\textbf{x}$ and $\textbf{y}$, $r = |\textbf{x}-\textbf{y}|$. In general, $\xi(r)$ also evolves with the redshift, but if one restricts the analysis to a given redshift bin, the correlation function in that redshift bin can be assumed to be constant. Since the dark matter is more abundant than the baryonic matter (which constitutes the `visible' galaxies, and intergalactic medium), the galaxies are expected to follow the gravitational potential well due to dark matter, and to a good approximation, at large scales, the 2PCF of galaxies $\xi_{gal}(r)$ will be related to the dark matter 2PCF $\xi_{DM}(r)$ via a factor $b_{gal}$ called `galaxy bias' as $\xi_{gal}(r) = b_{gal}^2 \, \xi_{DM}(r)$. In general, this bias, also known as `clustering bias' can be scale- and redshift-dependent \citep{Coles:2007be}.

\begin{figure}[ht]
\includegraphics[width=0.49\textwidth]{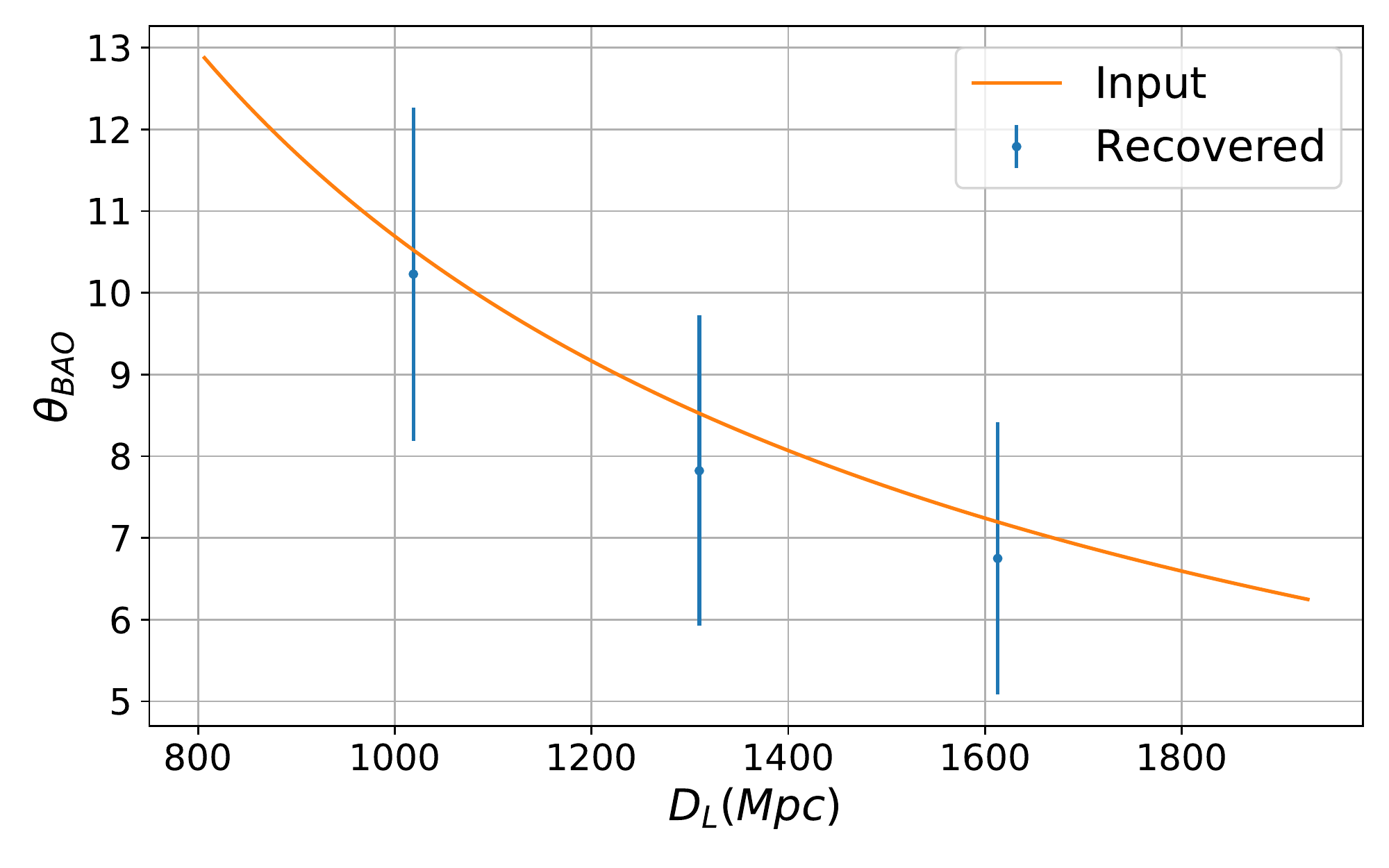}
\caption{The recovery of the BAO angular scale $\theta_{BAO}$ with BNS merger events (for a 3G detector network) centered around different values of $D_L$ in shells of about $\sim150 h^{-1} $ Mpc. The solid continuous curve represents the relation between $\theta_{BAO}$ and luminosity distance $D_L$ to the shell for $\Lambda$CDM model used in simulations using parameters from the \emph{Planck2018} results \citep{Planck:2018vyg}}
\label{fig:bao_recovery}
\end{figure}

\begin{table*}[ht]
    \caption{ The list of parametrized dark energy models considered in this work. $\Omega_{m0}$ and $\Omega_{r0}$ are present day density parameters for matter and radiation component respectively. $w, w_0$, and $w_a$ are related to the dark energy parametrization. The parameter $h$ is related to Hubble parameter $H_0$ as $h = \frac{H_0}{100~\text{km}~\text{s}^{-1}\text{Mpc}^{-1}}$
    .}
    
    \label{table:de_models}
    \begin{center}
    \begin{tabularx}{\textwidth}{cclc}
    \toprule \hline
    S. \\No. & Model & Hubble Equation  & parameters  \\
    
    \hline
    1. & $\Lambda$CDM & \centering \( h(z) = \sqrt{\Omega_{m_0}(1+z)^3+\Omega_{r_0}(1+z)^4 +1-\Omega_{m0}-\Omega_{r0}} \) &  $\Omega_{m0}, h$ \\
    \\
    2. & $w$CDM & \centering \( h(z) = \sqrt{\Omega_{m_0}(1+z)^3+\Omega_{r_0}(1+z)^4 +(1-\Omega_{m0}-\Omega_{r_0})(1+z)^{-3(1+w)}} \) &  $\Omega_{m0}, h,w$ \\
    \\
    3. & $w_0w_a$CDM & \centering \( h(z) = \sqrt{\Omega_{m0}(1+z)^3+\Omega_{r_0}(1+z)^4 +(1-\Omega_{m0}-\Omega_{r_0})(1+z)^{3(1+w_0+w_a)}\exp(-\frac{3w_a z}{1+z})} \) &  $\Omega_{m0}, h,w_0, w_a$ \\
 \\
    \hline
    \bottomrule
    \end{tabularx}
    \end{center}
\end{table*}

Apart from the galaxy bias, the 2PCF is also used to detect other LSS features, such as the BAO, from the distribution of matter. As $r_s$ can be considered a standard ruler, detecting the BAO peak at different redshifts provides an independent method to probe the cosmological parameters. Instead of using the three-dimensional correlation function $\xi(r)$, one can also use the two-point angular correlation function (2PACF) $\omega(\theta)$ by considering the galaxies in different redshift bins and projecting them along the radial direction in the shell, keeping in mind that the chosen shell should be small enough for the linear power spectrum $P(k,z)$ to remain constant in the redshift bin, i.e., $P(k,z) \sim P(k)$ for $z \in [z-dz/2, z+dz/2]$. The BAO scale $r_s$ is related to the angular scale $\theta_{BAO}$ and to the angular diameter distance $D_A$ as,
\begin{equation}
    \theta_{BAO}(z) = \frac{r_s}{(1+z)D_A(z)} \label{equation:bao_scale}
\end{equation}

By estimating the angular BAO scale $\theta_{BAO}(z)$ at a given redshift $z$, one can use the above relation to put constraints 
on the cosmological parameters which are embedded in the Hubble equation while calculating the angular diameter distance,
\begin{equation}
    D_A(z;\bm{\Theta}) = \frac{c}{H_0}\frac{1}{1+z}\int_0^z\frac{dz'}{h(z';\bm{\Theta})} \label{equation:angular_diameter_distance}
\end{equation}
\noindent
where $h(z; \bm{\Theta})$ is the normalized Hubble equation with parameters $\bm{\Theta}$ which depend on the cosmological model. For example, the normalized Hubble equation (at nearby redshifts) for the spatially flat $\Lambda$CDM model is,
\begin{equation}
    h(z) = \sqrt{\Omega_{m0}(1+z)^3+\Omega_{r0}(1+z)^4+1-\Omega_{m0}-\Omega_{r0}}, \label{equation:hubble_lcdm}
\end{equation}
\noindent
where $\Omega_{m0} = 3H_0^2\rho_m / (8\pi G)$ is called the density parameter for matter (which includes dark matter as well as baryonic matter) at present ($z=0$). $\Omega_{r0}$ is the density parameter corresponding to the radiation component. $\Omega_{d0}$ and $\Omega_{b0}$ represent density parameters corresponding to the dark matter and baryonic matter, respectively. It follows that $\Omega_{d0}+\Omega_{b0} = \Omega_{m0}$. Function $h(z)$ is called normalized Hubble parameter and is related to Hubble parameter $H(z)$ as $h(z)=\frac{H(z)}{H_0}$, where $H_0$ is the Hubble constant. In this study, we assume that the spatial curvature is zero, and we use the relation $\Omega_{m0}+\Omega_{r0}+\Omega_{\Lambda0} = 1$. 

In the spectroscopic surveys, the BAO scale appears in the line-of-sight direction and the transverse direction. In the line of sight direction, the Hubble parameter $H(z)$ can be measured as $c\Delta z/r_s$ where $\Delta z$ is the redshift range corresponding to the BAO scale. The corresponding Hubble distance at that redshift will be $D_H(z) = c/H(z)$. In the transverse direction, the BAO scale $r_s$ is related to the comoving angular diameter scale $D_M(z)$ to the angular BAO scale $\theta_{BAO}$ as $r_s = D_M(z) \theta_{BAO}$. The spectroscopic surveys provide the measurements of $D_H(z)/r_s$ and $D_M(z)/r_s$. This can be combined into a single quantity describing spherical averaged distance $D_V(z)/r_s$ where $D_V(z) \equiv [z D_M^2(z)D_H(z)]^{1/3}$ \citep{Giostri:2012ek,eBOSS:2020yzd}

\subsection{Probing the Large Scale Structures with Gravitational Waves}
As we expect the 3G detector network to provide a large number of well-localized GW merger events, the natural question arises: can we extend the similar methods as used for galaxies to probe the LSS with the distribution of GW observations using their localization volumes? Recent studies have shown that by cross-correlating localization volumes with galaxy catalogs, the LSS features such as the galaxy bias can be probed \citep{Mukherjee:2020hyn}. In this study, we are interested in probing the LSS purely with GW observations, without cross-correlation with galaxy catalogs.

The challenges in probing the LSS with just GW observations are twofold: i) the localization volumes obtained from the posteriors of GW events are currently very wide ($\mathcal{O}(10)-\mathcal{O}(100)$ Mpc for $D_L$ and \textbf{$\mathcal{O}(100)-\mathcal{O}(1000)$} square degrees for the sky localization) \citep{2018LRR....21....3A, 2022ApJ...924...54P}, which washes away most of the features in the LSS, and ii) the number of events which can be detected by current-generation GW detectors are not enough to probe the LSS. However, the planned 3G GW detectors such as the ET and CE are not only expected to have an order of magnitude better sensitivity compared to current detectors, but also expected to be more sensitive at low frequencies. This will allow the 3G detector network to detect enough events with precise enough localization volumes to make it possible to probe the LSS purely with GW events. It has been shown that with the 3G detector network, with 5-10 years of observation time, it will be possible to probe the galaxy bias solely from the GW events \citep{Vijaykumar:2020pzn}. Using the nearby BNS localization volumes ($z< 0.3$), the angular BAO scale $\theta_{BAO}$  can also be probed with the help of the 3G detector network \citep{Kumar:2021aog}.

\begin{figure*}[ht]
\includegraphics[width=0.95\textwidth]{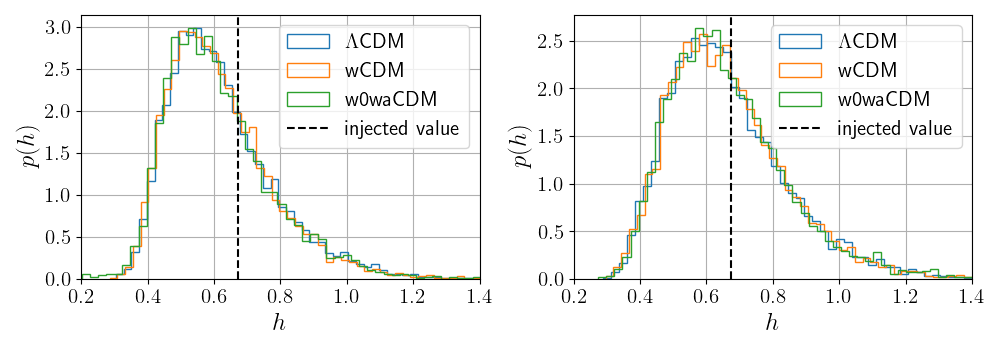}
\caption{The 1D marginalized posterior distribution of the Hubble parameter $H_0 = 100h~\mathrm{km}~\mathrm{s}^{-1}~\mathrm{Mpc}^{-1}$ using the angular BAO scale measurements $\theta_{BAO}$ from simulated GW events with the 3G detector network in three shells centered at $D_L \sim 1000~\mathrm{Mpc}$, $1300~\mathrm{Mpc}$, and $1600~\mathrm{Mpc}$. The injected value is shown by the dashed vertical line. The three cosmological models used here are described in table \ref{table:de_models}. Left panel shows GW-BAO+CMB constraints and right panel shows constraints from GW-BAO+CMB+EM counterpart.}
\label{fig:H0_GW_recovery}
\end{figure*}

The detection of the angular BAO scale $\theta_{BAO}$ at different redshifts from the GW localization volumes can be used as another cosmological probe. The localization volumes of the GW mergers can be divided into shells of luminosity distance $D_L(z)$ and $\theta_{BAO}(D_L)$ can be recovered \citep{Kumar:2021aog}. We can then use the measurement of the BAO scale $r_s$ from other surveys, such as CMB surveys, and use relation \ref{equation:bao_scale} to put constraints on the cosmological parameters. This gives us an independent approach to constrain cosmological parameters using combined GW-CMB data.

\section{Simulations and Results}
\label{sec:simulations}
We make use of the simulations done in \cite{Kumar:2021aog}, where we use publicly available code: {lognormal\_galaxies} \citep{Agrawal_2017} to create galaxy catalogs with the given correlation function $\xi(r)$ containing the BAO peak. These mock galaxy catalogs represent a realization of the Universe arising from the underlying density perturbations. 
These galaxies act as the host to the GW merger events. We then create a catalog of the BNS merger population consistent with the estimated merger rates obtained from the LVK analysis \citep{2021ApJ...913L...7A}. We use the network of 3G detectors containing an Einstein telescope \citep{Punturo_2010} in Europe, and two cosmic explorer detectors ~\citep{Reitze:2019iox,2021arXiv210909882E} located in the USA and Australia. In table \ref{table:detectors}, we show the detector configuration and location.

In this study, we restrict ourselves to the BNS sources between the redshift range ($0.2 \leq z \leq 0.3$) for the following reasons: 
\begin{itemize}
    \item The BNS merger rate is intrinsically higher ($10$ Gpc$^{-3}$ yr$^{-1} - 1700 $ Gpc$^{-3}$ yr$^{-1}$) compared to BBH merger rates ($\sim 17$ Gpc$^{-3}$ yr$^{-1} - 44$ Gpc$^{-3}$ yr$^{-1}$ at fiducial redshift $z=0.2$) \citep{LIGOScientific:2021psn}.
    \item Through simulations, we find that in the redshift range ($0.2 \leq z \leq 0.3$) we will have thousands of BNS events per year which are localized within a degree square in the sky \citep{Kumar:2021aog}. 
\end{itemize}
\noindent
Therefore, with $5-10$ years of accumulated data with 3G network, we will have enough highly localized BNS events which will enable us to calculate 2PACF $w(\theta)$ using the localization volumes from the posterior samples of detected BNS sources in various luminosity distance shells. We also show that using 2PACF, we will be able to detect the angular BAO scale $\theta_{BAO}$ in different luminosity distance shells by fitting for the BAO peak \citep{Kumar:2021aog}. We would like to emphasize that similar studies can be performed by other sources (like BBHs) if one can accumulate enough localized sources in a given $D_L$ shell. Although, as we go further, the localization volumes of the sources become larger.

In figure \ref{fig:bao_recovery}, we show the recovery of the angular BAO scale at different shells centered around $D_L \sim 1010$~Mpc, $1310$~Mpc, and $1620$~Mpc. We use these mock measurements of angular BAO scale $\theta_{BAO}$ (from simulations) as the input data. The difference between the angular BAO scale measurement from GW sources and that from Galaxy sources is that the former is done in $D_L$ space while later is done in redshift space. This makes the constraining power of the data sets complementary in terms of the set of parameters these data sets can constrain.
\begin{figure}[ht]
\includegraphics[width=0.49\textwidth]{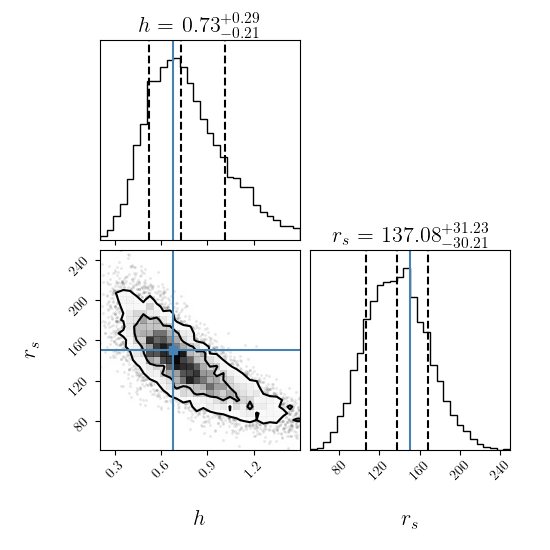}
\caption{The GW-BAO+EM counterpart constraints on the Hubble parameter $H_0 = 100h~\mathrm{km}~\mathrm{s}^{-1}~\mathrm{Mpc}^{-1}$ and comoving BAO scale $r_s$ [Mpc] for the $\Lambda$CDM model with BAO measurements from GW localization volumes (simulations). For the marginalized 1D posteriors, we show $68\%$ bounds with the median value. For 2D contours, we show $60\%$ and $90\%$ regions. The injected values are shown by the blue lines.}
\label{fig:lcdm_gw_rs}
\end{figure}

In the Bayesian framework, the posterior probability distribution $p(\bm{\Theta}|d,I)$ on the parameters $\bm{\Theta}$ given the data $d$, and any prior information $I$ is described as,
\begin{equation}
    p(\bm{\Theta}| d,I) = \frac{\mathcal{L}(d|\bm{\Theta},I)\pi(\bm{\Theta}|I)}{p(d|I)}, \label{equation:bayes}
\end{equation}
\noindent
where $\mathcal{L}(d|\bm{\Theta},I)$ is the likelihood function which represents the probability of the data given the parameters $\bm{\Theta}$ of the model. $\pi(\bm{\Theta}|I)$ is the prior probability distribution on the parameters, and $p(d|I)$ is called the `Bayesian evidence' or marginalized likelihood which acts as the normalization factor for the posterior distribution. We use the BAO measurements in different shells $\theta_{BAO}(D_L)$ as the data and define the likelihood function as, 
\begin{equation}
    \mathcal{L}(d|\bm{\Theta},I) \propto \exp( -\frac{1}{2} \sum_i \frac{(\theta_{BAO,i} - \theta_{BAO}(\bm{\Theta}))^2}{\sigma_i^2}) \label{equation:likelihood_gw_bao},
\end{equation}
\noindent

where $\Theta$ are the parameters describing the cosmological model. $\theta_{BAO,i}$ is the estimated angular BAO scale in the $i$-th shell corresponding to the effective luminosity distance $D_L^{i}$. $\sigma_i$ is the error associated with the measurement of the BAO peak in $i$-th shell. The reason for assuming this form of likelihood function is that through simulations, we find that recovered values of $\theta_{BAO}$ from $\sim$ 1000 catalogs we generated fits the Gaussian distribution around injected value. In general, the likelihood function \ref{equation:likelihood_gw_bao} will also have correlation between the different shells and the full covariance matrix is needed to be calculated and incorporated. But, in this study, we assume the shells to be independent and do not expect that to change the conclusions of the study. 

\begin{table*}
    \caption{The 3G detector network configuration (location, noise curves, and low frequency cutoff $f_\mathrm{low}$ ) used in the simulations done in \cite{Kumar:2021aog}. For the CE detector, subscript (2) represents (late) noise sensitivity curves and superscript (U, A) represents location of these detectors (USA, Australia). Similar detector configurations for CE and ET are taken in other studies such as in \cite{Nitz_2021}.
    }
    \label{table:detectors}
\begin{center}
\begin{tabular}{llllcc}
Abbreviation & Observatory  & $f_{\textrm{low}}$ &  Noise Curve & Latitude & Longitude \\ \hline
$C_2^U$ & Cosmic Explorer USA & 5.2 & CE2 & 40.8 & -113.8    \\
$C_2^A$ & Cosmic Explorer Australia & 5.2 & CE2 & -31.5 & 118.0   \\
$E$ & Einstein Telescope & 2 & ET-D Design & 43.6 & 10.5  \\

\end{tabular}
\end{center}
\end{table*}

\begin{figure}[ht]
\includegraphics[width=0.49\textwidth]{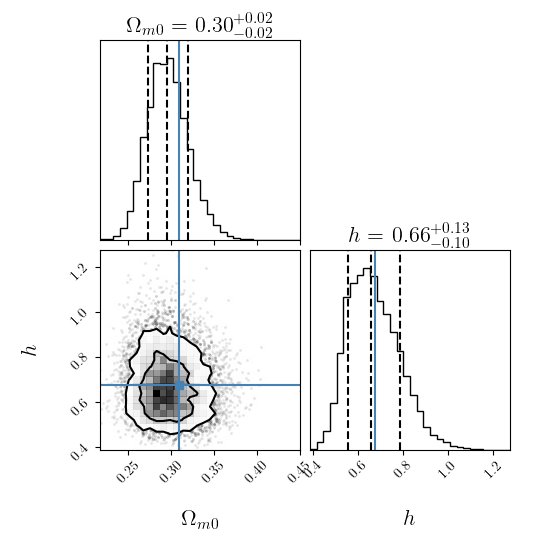}
\caption{The combined constraints (GW-BAO+SDSS-BAO+CMB) on the $\Lambda$CDM model with BAO measurements from GW localization volumes (from simulations) and CMB/BAO constraints from the galaxy surveys. For the marginalized 1D posteriors, we show $68\%$ bounds with the median value. For 2D contours, we show $60\%$ and $90\%$ regions. The injected values are shown by blue lines.}
\label{fig:lcdm_pe_recovery}
\end{figure}

We use standard $\Lambda$CDM model as reference model in our simulations. Other phenomenological dark energy models as described in table \ref{table:de_models} are also considered as recovery models, namely the $w$CDM model and CPL parametrization $w_0w_a$CDM model \citep{Chevallier:2000qy}. We make use of the publicly available implementation of nested sampling based sampler \textsc{Dynesty} \citep{speagle:2019} for parameter estimation. We use uniform prior on all the parameters in ranges $\Omega_{m0} \in [0.05,1.0]$, $\Omega_{r0}  \in [0.00001, 0.00015]$, $h \in [0.2,1.5]$, $w, w_0 \in [-2,0]$, and $w_a \in [-4,4]$. In our simulations, the BAO measurements from GW observations are conducted in luminosity distance shells of thickness 150 $h^{-1}$ Mpc and it is represented by the effective luminosity distance, $D_L^{eff}$, which is the midpoint of the $D_L$ shell. We demonstrate further that $\theta_{BAO}(D^{eff}_L)$ can be extracted at multiple $D^{eff}_L$ (see figure \ref{fig:bao_recovery}). To fit the overall shape of the curve from the $\theta_{BAO}-z$ relation (equation \ref{equation:bao_scale}), and account for the shell thickness, we treat the effective redshift of the shell, $z_{eff}$, as a free parameter corresponding to various $D_L$ shells. The likelihood function \ref{equation:likelihood_gw_bao} takes the form:
\begin{equation}
    \mathcal{L}(d|\bm{\Theta},I) \propto
    \exp( -\frac{1}{2} \sum_i \frac{(\theta_{BAO,i} - \theta_{BAO}(z_{eff}^i, \bm{\Theta}))^2}{\sigma_i^2}) 
    \label{equation:likelihood_gw_bao_z_prior},
\end{equation}
\noindent
where $\theta_{BAO,i}$ is the BAO angular scale measurement in $i$-th shell and 
\begin{equation}
    \theta_{BAO}(z_{eff}^i, \bm{\Theta}) = \frac{r_s}{(1+z_{eff}^i)D_A(z_{eff}^i, \bm{\Theta})},
\end{equation}
\noindent
is the angular BAO scale corresponding to the luminosity distance shell, and $\bm{\Theta}$ are the parameters of the cosmological model. We use the conditional uniform priors on $z_{eff}^i$ such that it includes the luminosity distance shell for reasonable cosmology models: $z_{eff}^i \in [z(D_{L, i}^{eff}-75, \bm{\Theta}),z(D_{L,i}^{eff}+75,\bm{\Theta})]$ for the three measurements of $\theta_{BAO}$ corresponding to the effective distance $D^{eff}_L  \sim 1010$~Mpc, $1310$~Mpc, and $1620$~Mpc. For a chosen cosmology model, $z(D_{L}, \bm{\Theta})$ represents the redshift corresponding to the luminosity distance $D_{L}$ and sampling parameters $\bm{\Theta}$.
We use two strategies for parameter $r_s$: 
\begin{enumerate}
    \item Calculate $r_s$ at given drag epoch $z_{d}$ by estimating the distance travelled by sound waves in early universe for a cosmology model and parameters: $\Omega_{r0}, \Omega_{m0}$, and $\Omega_{b0}$ \citep{Eisenstein:1997ik}. $\Omega_{b0}$ is set it to Planck 2018 value \citep{Planck:2018vyg}.  We call it GW-BAO+CMB constraints
    \item  We let the $r_s$ vary as free parameter and use the uniform prior in range $r_s \in [50, 250]$ Mpc. We call it GW-BAO constraints. 
\end{enumerate}

Additionally, we investigate a particular scenario in which we may have an EM counterpart associated with one of the BNS signals. In such cases, we can obtain precise values for $z_{eff}$ corresponding to the luminosity distance shell in which BNS with EM counterpart is detected. For this purpose, we use the luminosity distance shell corresponds to $D_L^{eff} \sim 1010$ Mpc from the simulation, and pivot it to fiducial observed redshift of $z_{eff}=0.2$. For all other shells, we use the conditional priors on $z_{eff}$ as described above. These constraints are referred to GW-BAO+CMB+EM counterpart constraints.

\begin{figure}[ht]
\includegraphics[width=0.49\textwidth]{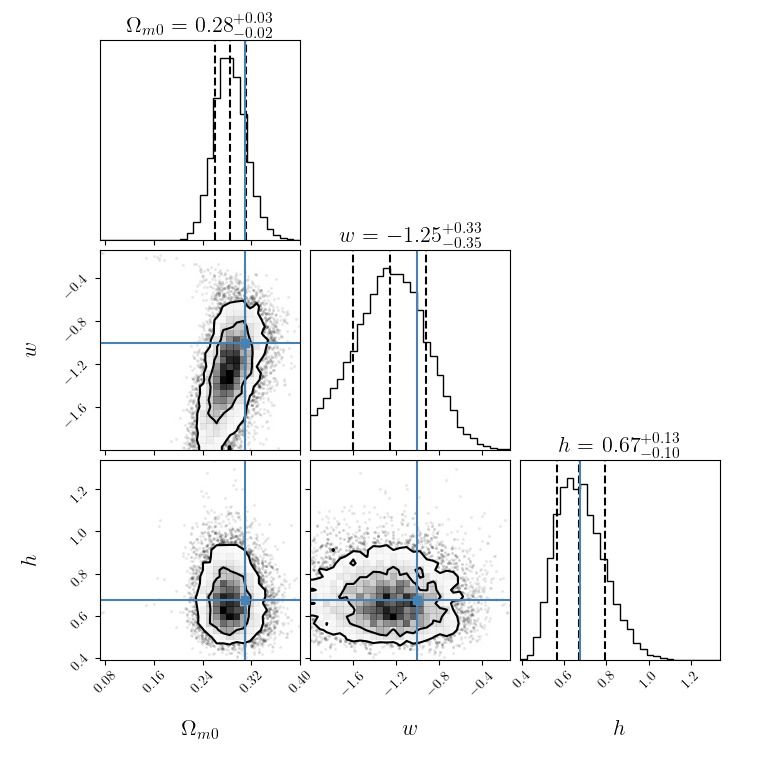}
\caption{The combined constraints (GW-BAO+SDSS-BAO+CMB) on the $w$CDM model with BAO measurements from GW localization volumes (from simulations) and CMB/BAO constraints from the galaxy surveys. For the marginalized 1D posteriors, we show $68\%$ bounds with the median value. For 2D contours, we show $60\%$ and $90\%$ regions. The injected values are shown by blue lines corresponding to the $\Lambda$CDM model with parameters consistent with the \emph{Planck2018} results \citep{Planck:2018vyg}.}
\label{fig:wcdm_pe_recovery}
\end{figure}

\begin{table*}[]
\Large
\centering
\caption{The constraints obtained on cosmological parameters corresponding to the models listed in table \ref{table:de_models}. For each parameter, 68\% (90\%) confidence intervals are reported. The GW-BAO constraints come from the BAO observation from the simulated GW BNS mergers for 3G detectors. For GW-BAO+CMB constraints, we calculate the comoving BAO scale $r_s$ from sampling cosmological parameters of the model. EM counterpart constraints indicate that one BNS merger event have an EM counterpart. SDSS-BAO+CMB constraints arise from the measurement of BAO features in SDSS data, and calculating the $r_s$ value. The entries with dashed lines indicate that the particular data do not constrain the given parameter i.e. the posterior obtained on these parameters just return the prior distribution. The empty entries indicate that the parameter is calculated from other sampling parameter and cosmology model. The ranges of the Uniform prior for each parameter are shown.}
\label{tab:pe_summary}
\resizebox{\textwidth}{!}{
\hspace{-5cm}
\begin{tabular}{@{}cccccccccc@{}}
\toprule
  \multicolumn{1}{c|}{Parameter} &
  \multicolumn{1}{c|}{Prior Range} &
  \multicolumn{1}{c|}{Model} &
  \multicolumn{1}{c|}{\begin{tabular}[c]{@{}c@{}}GW-BAO\\ +CMB \end{tabular}} &
  \multicolumn{1}{c|}{\begin{tabular}[c]{@{}c@{}}GW-BAO\\+EM counterpart \\($r_s$ varying) \end{tabular}} &
  \multicolumn{1}{c|}{\begin{tabular}[c]{@{}c@{}}GW-BAO\\ +CMB \\+EM counterpart \end{tabular}} &
  \multicolumn{1}{c|}{SDSS-BAO} &
  \multicolumn{1}{c|}{\begin{tabular}[c]{@{}c@{}}GW-BAO\\+SDSS-BAO\\+CMB \end{tabular}} &
  \multicolumn{1}{c|}{\begin{tabular}[c]{@{}c@{}}GW-BAO\\+SDSS-BAO\\+CMB \\+EM counterpart \end{tabular}} & \multicolumn{1}{c|}{\begin{tabular}[c]{@{}c@{}}GW-BAO\\+SDSS-BAO\\+CMB \\+EM counterpart\\ ($r_s$ varying)\end{tabular}} \\ \midrule &
  \\
\multirow{3}{*}{\begin{tabular}[c]{@{}c@{}}$H_0$\\ (km s$^{-1}$ Mpc$^{-1}$)\end{tabular}} &
  \multirow{3}{*}{$\mathcal{U}$(20,150)} &
  $\Lambda$CDM &
  $59.4^{+18.0}_{-12.1} (59.4^{+ 33.9}_{-17.7})$ &
  $72.9^{+28.5}_{-20.8} (72.9^{+ 49.8}_{-31.6})$ &
  $63.5^{+20.0}_{-14.3} (63.5^{+ 37.0}_{-21.3})$ &
  -- &
  $65.9^{+12.8}_{-10.0} (65.9^{+ 21.4}_{-14.8})$ &
  $68.8^{+15.8}_{-12.2} (68.8^{+ 27.6}_{-18.5})$ &
  $74.3^{+28.5}_{-20.4} (74.3^{+ 50.2}_{-31.3})$ \\ 
 &
   &
  $w$CDM &
  $59.0^{+17.4}_{-12.0} (59.0^{+ 32.2}_{-17.7})$ &
  $73.3^{+28.5}_{-20.9} (73.3^{+ 51.8}_{-32.2})$ &
  $63.9^{+19.3}_{-14.5} (63.9^{+ 35.6}_{-21.3})$ &
  -- &
  $66.9^{+12.6}_{-10.3} (66.9^{+ 21.7}_{-15.4})$ &
  $69.5^{+15.7}_{-12.4} (69.5^{+ 27.9}_{-18.9})$ &
  $75.8^{+29.0}_{-21.5} (75.8^{+ 51.1}_{-32.2})$ \\ 
 &
   &
  $w_0w_a$CDM &
  $58.7^{+18.1}_{-11.9} (58.7^{+ 33.9}_{-17.7})$ &
  $72.9^{+29.3}_{-20.9} (72.9^{+ 51.9}_{-31.6})$ &
  $63.4^{+19.3}_{-13.9} (63.4^{+ 36.3}_{-20.6})$ &
  -- &
  $65.9^{+12.7}_{-10.2} (65.9^{+ 21.8}_{-15.5})$ &
  $68.3^{+15.8}_{-12.0} (68.3^{+ 27.9}_{-18.3})$ &  $75.5^{+28.7}_{-21.7} (75.5^{+ 51.0}_{-32.5})$ \\ \cmidrule(r){1-3}
  \\
\multirow{3}{*}{$\Omega_{m0}$} &
  \multirow{3}{*}{$\mathcal{U}$(0.05, 1)} &
  $\Lambda$CDM &
  -- &
  -- &
  -- &
  $0.30^{+0.02}_{-0.02} (0.30^{+ 0.04}_{-0.04})$ &
  $0.30^{+0.02}_{-0.02} (0.30^{+ 0.04}_{-0.04})$ &
  $0.29^{+0.03}_{-0.02} (0.29^{+ 0.04}_{-0.04})$ &   $0.30^{+0.03}_{-0.02} (0.30^{+ 0.04}_{-0.04})$ \\ 
 &
   &
  $w$CDM &
  -- &
  -- &
  -- &
  $0.28^{+0.03}_{-0.03} (0.28^{+ 0.05}_{-0.04})$ &
  $0.28^{+0.03}_{-0.02} (0.28^{+ 0.05}_{-0.04})$ &
  $0.29^{+0.03}_{-0.03} (0.29^{+ 0.05}_{-0.04})$ & 
  $0.28^{+0.03}_{-0.03} (0.28^{+ 0.05}_{-0.04})$ \\ 
 &
   &
  $w_0w_a$CDM &
  -- &
  -- &
  -- &
  $0.29^{+0.04}_{-0.04} (0.29^{+ 0.06}_{-0.07})$ &
  $0.29^{+0.04}_{-0.03} (0.29^{+ 0.06}_{-0.06})$ &
  $0.30^{+0.04}_{-0.03} (0.30^{+ 0.06}_{-0.05})$ & 
  $0.29^{+0.04}_{-0.04} (0.29^{+ 0.06}_{-0.07})$ \\ \cmidrule(r){1-3}
  \\
\multirow{3}{*}{\begin{tabular}[c]{@{}c@{}}$r_s$\\ (Mpc)\end{tabular}} &
  \multirow{3}{*}{$\mathcal{U}$(50, 250)} &
  $\Lambda$CDM &
   &
  $137.1^{+31.2}_{-30.2} (137.1^{+ 52.8}_{-46.5})$ &
   & 
   &
   &
   &
   $137.2^{+32.0}_{-29.5} (137.2^{+ 53.4}_{-45.1})$ \\ 
 &
   &
  $w$CDM &
   &
  $136.8^{+32.1}_{-29.8} (136.8^{+ 54.2}_{-47.6})$ &
   &
   &
   &
   &
   $138.4^{+31.7}_{-29.8} (138.4^{+ 52.0}_{-45.8})$ \\ 
 &
   &
  $w_0w_a$CDM &
   &
  $137.3^{+31.3}_{-31.4} (137.3^{+ 52.3}_{-47.7})$ &
   &
   &
   &
   &
   $138.0^{+32.1}_{-30.4} (138.0^{+ 53.0}_{-46.4})$ \\ \cmidrule(r){1-3}
   \\
$w$ &
  $\mathcal{U}$(-2,0) &
  $w$CDM &
  -- &
  -- &
  -- &
 $-1.25^{+0.34}_{-0.35} (-1.25^{+ 0.54}_{-0.56})$ & 
 $-1.25^{+0.33}_{-0.35} (-1.25^{+ 0.53}_{-0.57})$ & 
 $-1.23^{+0.33}_{-0.36} (-1.23^{+ 0.53}_{-0.56})$ & 
 $-1.25^{+0.33}_{-0.36} (-1.25^{+ 0.53}_{-0.57})$   \\ \cmidrule(r){1-3}
 \\
$w_0$ &
  $\mathcal{U}$(-2, 0) &
  $w_0w_a$CDM &
  -- &
  -- &
  -- &
  $-1.05^{+0.46}_{-0.47} (-1.05^{+ 0.73}_{-0.75})$ &
  $-1.05^{+0.47}_{-0.47} (-1.05^{+ 0.73}_{-0.73})$ &
  $-1.02^{+0.47}_{-0.49} (-1.02^{+ 0.72}_{-0.76})$ &
  $-1.06^{+0.47}_{-0.48} (-1.06^{+ 0.73}_{-0.74})$ \\ \cmidrule(r){1-3}
  \\
$w_a$ &
  $\mathcal{U}$(-4, 4) &
  $w_0w_a$CDM &
  -- &
  -- &
  -- &
  $-1.31^{+2.05}_{-1.86} (-1.31^{+ 2.79}_{-2.42})$ &
  $-1.43^{+2.05}_{-1.76} (-1.43^{+ 2.78}_{-2.30})$ &
  $-1.45^{+1.95}_{-1.77} (-1.45^{+ 2.71}_{-2.30})$ &
  $-1.28^{+2.02}_{-1.85} (-1.28^{+ 2.75}_{-2.45})$ \\ 
  \\
  \hline
\end{tabular}
}
\end{table*}

The Bayesian analysis with $\theta_{BAO}(D_L)$ measurements from the simulations provide constraints only on the Hubble parameter $H_0$. Other cosmological parameters returns the uniform prior distribution indicating that data lacks the power to constrain these parameters. In figure \ref{fig:H0_GW_recovery}, we show the recovery of the Hubble parameter for different cosmological models. It turns out that with GW-BAO+CMB data it is possible to constrain the Hubble parameter $H_0$. The inferred value of $H_0$ ($90\%$ CL) for different models considered here are: i) $\Lambda$CDM model: $H_0 = 59.4^{+ 33.9}_{-17.7}~\mathrm{km}~\mathrm{s}^{-1}~\mathrm{Mpc}^{-1}$, ii) $w$CDM model: $H_0 = 59.0^{+ 32.2}_{-17.7}~\mathrm{km}~\mathrm{s}^{-1}~\mathrm{Mpc}^{-1}$, and iii) $w_0w_a$CDM model: $H_0 = 58.7^{+ 33.9}_{-17.7}~\mathrm{km}~\mathrm{s}^{-1}~\mathrm{Mpc}^{-1}$. For GW-BAO+CMB+EM constraints, the inferred $H_0$ values for different models turns out to be: i) $\Lambda$CDM model: $H_0 = 63.5^{+ 37.0}_{-21.3}~\mathrm{km}~\mathrm{s}^{-1}~\mathrm{Mpc}^{-1}$, ii) $w$CDM model: $H_0 = 63.9^{+ 35.6}_{-21.3}~\mathrm{km}~\mathrm{s}^{-1}~\mathrm{Mpc}^{-1}$, and iii) $w_0w_a$CDM model: $H_0 = 63.4^{+ 36.3}_{-20.6}~\mathrm{km}~\mathrm{s}^{-1}~\mathrm{Mpc}^{-1}$. The injected value for the $H_0$ is $67.04~\mathrm{km}~\mathrm{s}^{-1}~\mathrm{Mpc}^{-1}$.

For GW-BAO + EM counterpart constraints, where $r_s$ is treated as free parameter, the results are shown in figure \ref{fig:lcdm_gw_rs}. In this case, we are able to constrain two parameters: $H_0$ and $r_s$. As expected, the constraints on the $H_0$ are wider compared to the case where acoustic length scale $r_s$ was not treated as free parameter.

BAO measurements alone from galaxy surveys provide the constraints on the $\Omega_{m0}$ but not on the Hubble parameter $H_0$. Therefore, combining the GW-BAO measurements and galaxy-BAO measurements shall provide the combined constraints on parameters $\Omega_{m0}$ and $H_0$. We expect that current and future spectroscopic surveys such as SDSS \citep{2023arXiv230107688A}, Euclid \citep{euclidmission}, Vera C. Rubin Observatory (LSST) \citep{LSST:2008ijt} will provide more robust BAO measurements by the time 3G detectors are operational. However, in this study, we use current galaxy-BAO measurements and combine them with projected GW-BAO measurements from the simulations done for 3G GW detectors to get conservative estimates of constraining power of the combination of the data. We use the current BAO  measurements for the angular scales: clustering measurements on $D_M(z)/r_s$ and $D_V(z)/r_s$ at various redshifts as compiled in the data from SDSS, SDSS-II, BOSS, and eBOSS \citep{eBOSS:2020yzd}. We call the constraints from Galaxy-BAO likelihood functions to be SDSS-BAO+CMB constraints. We combined data sets by multiplying the GW-BAO+CMB likelihood and SDSS-BAO+CMB likelihood.

\begin{center}
\begin{figure*}
\includegraphics[width=0.9\textwidth]{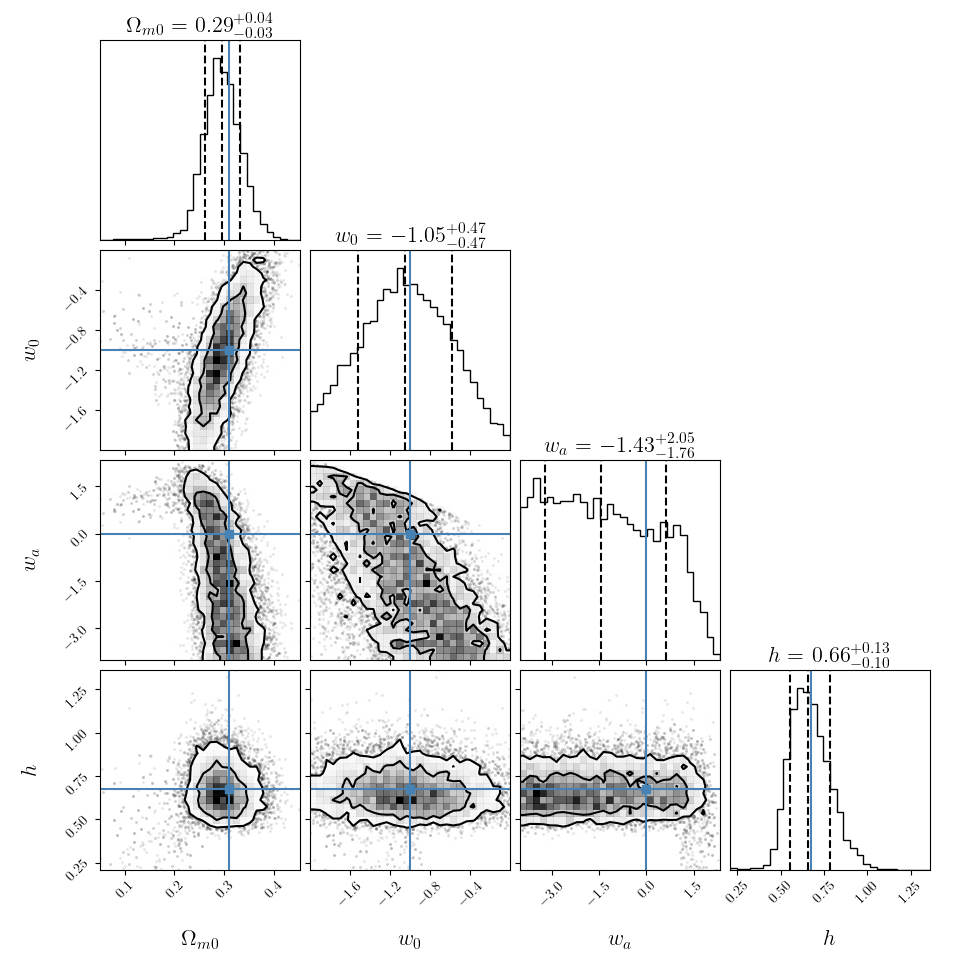}
\caption{The combined constraints (GW-BAO+SDSS-BAO+CMB) on the $w_0w_a$CDM model with BAO measurements from GW localization volumes (from simulations) and CMB/BAO constraints from the galaxy surveys. For the marginalized 1D posteriors, we show $68\%$ bounds with the median value. For 2D contours, we show $60\%$ and $90\%$ regions. The injected values are shown by blue lines corresponding to the $\Lambda$CDM model with parameters consistent with the \emph{Planck2018} results \citep{Planck:2018vyg}.}
\label{fig:w0wacdm_pe_recovery}
\end{figure*}
\end{center}

In figures \ref{fig:lcdm_pe_recovery},\ref{fig:wcdm_pe_recovery}, and \ref{fig:w0wacdm_pe_recovery}, we show the combined constraints (GW-BAO+SDSS-BAO+CMB) on the parameters of the models $\Lambda$CDM, $w$CDM, and $w_0w_a$CDM respectively. In table \ref{tab:pe_summary}, we collect results from the constraints obtained on various cosmological parameters with different combinations of data sets. We observe that i) GW-BAO+CMB data alone can constrain the Hubble parameter but not other cosmological parameters. ii) Spectroscopic BAO measurements alone can not constrain the Hubble parameter but they can constrain other cosmological parameters such as density parameter $\Omega_{m0}$, and dark energy parameters. iii) These two data sets are complementary to each other and hence combining them will allow us to constrain the cosmological models from BAO measurements alone, and iv) the constraints on $H_0$ are relatively weaker if we allow the parameter $r_s$ to vary.
 
Although, as an example, we show here only a few  selected parametrized DE models, this could also be applied to study other DE models, such as canonical and non-canonical scalar field models \citep{Wetterich:1987fm, Ratra:1987rm, Peebles:1987ek, Turner:1997npq, Caldwell:1997ii, Zlatev:1998tr, Bagla:2002yn} , Galileon models \citep{Nicolis:2008in, Ali:2010gr, Gannouji:2010au}, and other models of cosmology based on  modified gravity theories \citep{Clifton_2012}.\\

\section{Summary}
\label{sec:summary}
The future of GW cosmology looks bright as the growing catalog of GW mergers provides us an independent probe of the Universe apart from the traditional electromagnetic window. The independent probes offers us not only additional opportunities to constrain the cosmological parameters, they might also help to resolve the possible tension between various data sets such as the so called Hubble tension between current CMB data at high redshift \citep{Planck:2018vyg} and SNe data from the low redshifts \citep{Riess:2019cxk}. 
With the current catalog of GW events, the localization volumes (from BBHs) can be used along with galaxy catalogs using cross correlation techniques to constrain the Hubble constant $H_0$ \citep{LIGOScientific:2019zcs}. In case of BNS events which have electromagnetic counterparts (e.g. GW170817), more stringent constraints on $H_0$ can be put because of the precise redshift information \citep{LIGOScientific:2017adf}. In the future, we expect these constraints to become stringent with more GW merger observations as we expect to detect EM counterparts for a small fraction of events (such as nearby BNS/NSBH mergers). 

The third generation of GW detectors such as ET and CE are expected to be order of magnitude more sensitive than current generation detectors, and will be able to probe lower frequencies upto few Hz \citep{Reitze:2019iox, Sathyaprakash:2012jk}. It will enable them to detect thousands of GW mergers with precise enough localization to probe the large scale structures of the Universe using their localization volumes solely from the GW merger observations \citep{Vijaykumar:2020pzn}. We should be able to probe the LSS features such as galaxy bias, and BAO peak by measuring the 2PCF from the localization volumes \citep{Kumar:2021aog, Vijaykumar:2020pzn}. In this study we show that, with 3G detector network, by tracing the angular BAO scale from GW mergers $\theta_{BAO}(D_L)$ at various luminosity distance bins, we can put constraints on the cosmological parameters such as the Hubble constant $H_0$ (for $\Lambda$CDM model) with $90\%$ credible intervals $H_0=59.4^{+ 33.9}_{-17.7} ~\mathrm{km}~\mathrm{s}^{-1}~\mathrm{Mpc}^{-1}$.  We show that the constraints on the cosmological parameters from GW-BAO data are complementary to the constraints obtained from galaxy-BAO measurements. Therefore, when these data sets are combined, it will enable us to constrain the parameters of various dark energy models. In this study, as a proof of concept, we combine the expected BAO constraints from GW mergers from the 3G detector network along with the BAO measurements from current spectroscopic surveys, though we expect the future spectroscopic surveys to be outperforming the current generation of galaxy surveys. Therefore, the results presented in this study are the conservative estimates. We would also like to emphasize that, this is not a unique method to put constraints on cosmological parameters as more stringent constraints can be provided by various combination of data from other cosmological surveys e.g. CMB, type Ia supernovae, etc. However, it will still provide an independent probe of cosmology which can be combined with the available data from future galaxy surveys to put tighter constraints and in the best case, help in resolving the tension between the competing data sets, if any. This study adds to the science case of 3G detectors and build on the previous studies on probing LSS with 3G detectors network \citep{Kumar:2021aog, Vijaykumar:2020pzn}. 
\acknowledgments
We acknowledge the Max Planck Gesellschaft. We thank the computing team from AEI Hannover for their significant technical support. SK thanks Xisco Jim\'enez Forteza and Pierre Mourier for going through the manuscript and providing useful comments. SK also thanks Aditya Vijaykumar and Anjan Sen for useful discussions. SK thanks the anonymous referee for critical inputs and improving the manuscript. ATLAS cluster at AEI Hannover was used to perform all the computational work done in this study.

\bibliography{references}

\begin{thebibliography}{}
\expandafter\ifx\csname natexlab\endcsname\relax\def\natexlab#1{#1}\fi
\providecommand{\url}[1]{\href{#1}{#1}}
\providecommand{\dodoi}[1]{doi:~\href{http://doi.org/#1}{\nolinkurl{#1}}}
\providecommand{\doeprint}[1]{\href{http://ascl.net/#1}{\nolinkurl{http://ascl.net/#1}}}
\providecommand{\doarXiv}[1]{\href{https://arxiv.org/abs/#1}{\nolinkurl{https://arxiv.org/abs/#1}}}

\bibitem[{Aasi {et~al.}(2015)}]{LIGOScientific:2014pky}
Aasi, J., {et~al.} 2015, Class. Quant. Grav., 32, 074001,
  \dodoi{10.1088/0264-9381/32/7/074001}

\bibitem[{Abbott {et~al.}(2016)}]{LIGOScientific:2016vbw}
Abbott, B.~P., {et~al.} 2016, Phys. Rev. D, 93, 122003,
  \dodoi{10.1103/PhysRevD.93.122003}

\bibitem[{Abbott {et~al.}(2017{\natexlab{a}})Abbott, Abbott, Abbott, Acernese,
  Ackley, Adams, Adams, Addesso, Adhikari, Adya, Affeldt, Afrough, Agarwal,
  Agathos, Agatsuma, Aggarwal, Aguiar, Aiello, Ain, Ajith, Allen, Allen,
  Allocca, Altin, Amato, Ananyeva, Anderson, Anderson, Angelova, Antier,
  Appert, Arai, Araya, Areeda, Arnaud, Arun, Ascenzi, Ashton, Ast, Aston,
  Astone, Atallah, Aufmuth, Aulbert, AultONeal, Austin, Avila-Alvarez, Babak,
  Bacon, Bader, Bae, Bailes, Baker, Baldaccini, Ballardin, Ballmer, Banagiri,
  Barayoga, Barclay, Barish, Barker, Barkett, Barone, Barr, Barsotti,
  Barsuglia, Barta, Barthelmy, Bartlett, Bartos, Bassiri, Basti, Batch, Bawaj,
  Bayley, Bazzan, B\'ecsy, Beer, Bejger, Belahcene, Bell, Berger, Bergmann,
  Bernuzzi, Bero, Berry, Bersanetti, Bertolini, Betzwieser, Bhagwat, Bhandare,
  Bilenko, Billingsley, Billman, Birch, Birney, Birnholtz, Biscans, Biscoveanu,
  Bisht, Bitossi, Biwer, Bizouard, Blackburn, Blackman, Blair, Blair, Blair,
  Bloemen, Bock, Bode, Boer, Bogaert, Bohe, Bondu, Bonilla, Bonnand, Boom,
  Bork, Boschi, Bose, Bossie, Bouffanais, Bozzi, Bradaschia, Brady, Branchesi,
  Brau, Briant, Brillet, Brinkmann, Brisson, Brockill, Broida, Brooks, Brown,
  Brown, Brunett, Buchanan, Buikema, Bulik, Bulten, Buonanno, Buskulic, Buy,
  Byer, Cabero, Cadonati, Cagnoli, Cahillane, Calder\'on~Bustillo, Callister,
  Calloni, Camp, Canepa, Canizares, Cannon, Cao, Cao, Capano, Capocasa,
  Carbognani, Caride, Carney, Carullo, Casanueva~Diaz, Casentini, Caudill,
  Cavagli\`a, Cavalier, Cavalieri, Cella, Cepeda, Cerd\'a-Dur\'an, Cerretani,
  Cesarini, Chamberlin, Chan, Chao, Charlton, Chase, Chassande-Mottin,
  Chatterjee, Chatziioannou, Cheeseboro, Chen, Chen, Chen, Cheng, Chia,
  Chincarini, Chiummo, Chmiel, Cho, Cho, Chow, Christensen, Chu, Chua, Chua,
  Chung, Chung, Ciani, Ciolfi, Cirelli, Cirone, Clara, Clark, Clearwater,
  Cleva, Cocchieri, Coccia, Cohadon, Cohen, Colla, Collette, Cominsky,
  Constancio, Conti, Cooper, Corban, Corbitt, Cordero-Carri\'on, Corley,
  Cornish, Corsi, Cortese, Costa, Coughlin, Coughlin, Coulon, Countryman,
  Couvares, Covas, Cowan, Coward, Cowart, Coyne, Coyne, Creighton, Creighton,
  Cripe, Crowder, Cullen, Cumming, Cunningham, Cuoco, Dal~Canton, D\'alya,
  Danilishin, D'Antonio, Danzmann, Dasgupta, Da~Silva~Costa, Dattilo, Dave,
  Davier, Davis, Daw, Day, De, DeBra, Degallaix, De~Laurentis, Del\'eglise,
  Del~Pozzo, Demos, Denker, Dent, De~Pietri, Dergachev, De~Rosa, DeRosa,
  De~Rossi, DeSalvo, de~Varona, Devenson, Dhurandhar, D\'{\i}az, Dietrich,
  Di~Fiore, Di~Giovanni, Di~Girolamo, Di~Lieto, Di~Pace, Di~Palma, Di~Renzo,
  Doctor, Dolique, Donovan, Dooley, Doravari, Dorrington, Douglas,
  Dovale~\'Alvarez, Downes, Drago, Dreissigacker, Driggers, Du, Ducrot, Dudi,
  Dupej, Dwyer, Edo, Edwards, Effler, Eggenstein, Ehrens, Eichholz, Eikenberry,
  Eisenstein, Essick, Estevez, Etienne, Etzel, Evans, Evans, Factourovich,
  Fafone, Fair, Fairhurst, Fan, Farinon, Farr, Farr, Fauchon-Jones, Favata,
  Fays, Fee, Fehrmann, Feicht, Fejer, Fernandez-Galiana, Ferrante, Ferreira,
  Ferrini, Fidecaro, Finstad, Fiori, Fiorucci, Fishbach, Fisher, Fitz-Axen,
  Flaminio, Fletcher, Fong, Font, Forsyth, Forsyth, Fournier, Frasca, Frasconi,
  Frei, Freise, Frey, Frey, Fries, Fritschel, Frolov, Fulda, Fyffe, Gabbard,
  Gadre, Gaebel, Gair, Gammaitoni, Ganija, Gaonkar, Garcia-Quiros, Garufi,
  Gateley, Gaudio, Gaur, Gayathri, Gehrels, Gemme, Genin, Gennai, George,
  George, Gergely, Germain, Ghonge, Ghosh, Ghosh, Ghosh, Giaime, Giardina,
  Giazotto, Gill, Glover, Goetz, Goetz, Gomes, Goncharov, Gonz\'alez,
  Gonzalez~Castro, Gopakumar, Gorodetsky, Gossan, Gosselin, Gouaty, Grado,
  Graef, Granata, Grant, Gras, Gray, Greco, Green, Gretarsson, Groot, Grote,
  Grunewald, Gruning, Guidi, Guo, Gupta, Gupta, Gushwa, Gustafson, Gustafson,
  Halim, Hall, Hall, Hamilton, Hammond, Haney, Hanke, Hanks, Hanna, Hannam,
  Hannuksela, Hanson, Hardwick, Harms, Harry, Harry, Hart, Haster, Haughian,
  Healy, Heidmann, Heintze, Heitmann, Hello, Hemming, Hendry, Heng, Hennig,
  Heptonstall, Heurs, Hild, Hinderer, Ho, Hoak, Hofman, Holt, Holz, Hopkins,
  Horst, Hough, Houston, Howell, Hreibi, Hu, Huerta, Huet, Hughey, Husa,
  Huttner, Huynh-Dinh, Indik, Inta, Intini, Isa, Isac, Isi, Iyer, Izumi,
  Jacqmin, Jani, Jaranowski, Jawahar, Jim\'enez-Forteza, Johnson,
  Johnson-McDaniel, Jones, Jones, Jonker, Ju, Junker, Kalaghatgi, Kalogera,
  Kamai, Kandhasamy, Kang, Kanner, Kapadia, Karki, Karvinen, Kasprzack,
  Kastaun, Katolik, Katsavounidis, Katzman, Kaufer, Kawabe, K\'ef\'elian,
  Keitel, Kemball, Kennedy, Kent, Key, Khalili, Khan, Khan, Khan, Khazanov,
  Kijbunchoo, Kim, Kim, Kim, Kim, Kim, Kim, Kimbrell, King, King,
  Kinley-Hanlon, Kirchhoff, Kissel, Kleybolte, Klimenko, Knowles, Koch,
  Koehlenbeck, Koley, Kondrashov, Kontos, Korobko, Korth, Kowalska, Kozak,
  Kr\"amer, Kringel, Krishnan, Kr\'olak, Kuehn, Kumar, Kumar, Kumar, Kuo,
  Kutynia, Kwang, Lackey, Lai, Landry, Lang, Lange, Lantz, Lanza, Larson,
  Lartaux-Vollard, Lasky, Laxen, Lazzarini, Lazzaro, Leaci, Leavey, Lee, Lee,
  Lee, Lee, Lee, Lehmann, Lenon, Leon, Leonardi, Leroy, Letendre, Levin, Li,
  Linker, Littenberg, Liu, Liu, Lo, Lockerbie, London, Lord, Lorenzini,
  Loriette, Lormand, Losurdo, Lough, Lousto, Lovelace, L\"uck, Lumaca,
  Lundgren, Lynch, Ma, Macas, Macfoy, Machenschalk, MacInnis, Macleod, Maga\~na
  Hernandez, Maga\~na Sandoval, Maga\~na Zertuche, Magee, Majorana, Maksimovic,
  Man, Mandic, Mangano, Mansell, Manske, Mantovani, Marchesoni, Marion,
  M\'arka, M\'arka, Markakis, Markosyan, Markowitz, Maros, Marquina, Marsh,
  Martelli, Martellini, Martin, Martin, Martynov, Marx, Mason, Massera,
  Masserot, Massinger, Masso-Reid, Mastrogiovanni, Matas, Matichard, Matone,
  Mavalvala, Mazumder, McCarthy, McClelland, McCormick, McCuller, McGuire,
  McIntyre, McIver, McManus, McNeill, McRae, McWilliams, Meacher, Meadors,
  Mehmet, Meidam, Mejuto-Villa, Melatos, Mendell, Mercer, Merilh, Merzougui,
  Meshkov, Messenger, Messick, Metzdorff, Meyers, Miao, Michel, Middleton,
  Mikhailov, Milano, Miller, Miller, Miller, Millhouse, Milovich-Goff,
  Minazzoli, Minenkov, Ming, Mishra, Mitra, Mitrofanov, Mitselmakher,
  Mittleman, Moffa, Moggi, Mogushi, Mohan, Mohapatra, Molina, Montani, Moore,
  Moraru, Moreno, Morisaki, Morriss, Mours, Mow-Lowry, Mueller, Muir,
  Mukherjee, Mukherjee, Mukherjee, Mukund, Mullavey, Munch, Mu\~niz, Muratore,
  Murray, Nagar, Napier, Nardecchia, Naticchioni, Nayak, Neilson, Nelemans,
  Nelson, Nery, Neunzert, Nevin, Newport, Newton, Ng, Nguyen, Nguyen, Nichols,
  Nielsen, Nissanke, Nitz, Noack, Nocera, Nolting, North, Nuttall, Oberling,
  O'Dea, Ogin, Oh, Oh, Ohme, Okada, Oliver, Oppermann, Oram, O'Reilly,
  Ormiston, Ortega, O'Shaughnessy, Ossokine, Ottaway, Overmier, Owen, Pace,
  Page, Page, Pai, Pai, Palamos, Palashov, Palomba, Pal-Singh, Pan, Pan, Pang,
  Pang, Pankow, Pannarale, Pant, Paoletti, Paoli, Papa, Parida, Parker,
  Pascucci, Pasqualetti, Passaquieti, Passuello, Patil, Patricelli, Pearlstone,
  Pedraza, Pedurand, Pekowsky, Pele, Penn, Perez, Perreca, Perri, Pfeiffer,
  Phelps, Piccinni, Pichot, Piergiovanni, Pierro, Pillant, Pinard, Pinto,
  Pirello, Pitkin, Poe, Poggiani, Popolizio, Porter, Post, Powell, Prasad,
  Pratt, Pratten, Predoi, Prestegard, Prijatelj, Principe, Privitera, Prix,
  Prodi, Prokhorov, Puncken, Punturo, Puppo, P\"urrer, Qi, Quetschke, Quintero,
  Quitzow-James, Raab, Rabeling, Radkins, Raffai, Raja, Rajan, Rajbhandari,
  Rakhmanov, Ramirez, Ramos-Buades, Rapagnani, Raymond, Razzano, Read,
  Regimbau, Rei, Reid, Reitze, Ren, Reyes, Ricci, Ricker, Rieger, Riles, Rizzo,
  Robertson, Robie, Robinet, Rocchi, Rolland, Rollins, Roma, Romano, Romano,
  Romel, Romie, Rosi\ifmmode~\acute{n}\else \'{n}\fi{}ska, Ross, Rowan,
  R\"udiger, Ruggi, Rutins, Ryan, Sachdev, Sadecki, Sadeghian, Sakellariadou,
  Salconi, Saleem, Salemi, Samajdar, Sammut, Sampson, Sanchez, Sanchez,
  Sanchis-Gual, Sandberg, Sanders, Sassolas, Sathyaprakash, Saulson, Sauter,
  Savage, Sawadsky, Schale, Scheel, Scheuer, Schmidt, Schmidt, Schnabel,
  Schofield, Sch\"onbeck, Schreiber, Schuette, Schulte, Schutz, Schwalbe,
  Scott, Scott, Seidel, Sellers, Sengupta, Sentenac, Sequino, Sergeev,
  Shaddock, Shaffer, Shah, Shahriar, Shaner, Shao, Shapiro, Shawhan, Sheperd,
  Shoemaker, Shoemaker, Siellez, Siemens, Sieniawska, Sigg, Silva, Singer,
  Singh, Singhal, Sintes, Slagmolen, Smith, Smith, Smith, Somala, Son,
  Sonnenberg, Sorazu, Sorrentino, Souradeep, Spencer, Srivastava, Staats,
  Staley, Steinke, Steinlechner, Steinlechner, Steinmeyer, Stevenson, Stone,
  Stops, Strain, Stratta, Strigin, Strunk, Sturani, Stuver, Summerscales, Sun,
  Sunil, Suresh, Sutton, Swinkels, Szczepa\ifmmode~\acute{n}\else
  \'{n}\fi{}czyk, Tacca, Tait, Talbot, Talukder, Tanner, T\'apai, Taracchini,
  Tasson, Taylor, Taylor, Tewari, Theeg, Thies, Thomas, Thomas, Thomas, Thorne,
  Thorne, Thrane, Tiwari, Tiwari, Tokmakov, Toland, Tonelli, Tornasi,
  Torres-Forn\'e, Torrie, T\"oyr\"a, Travasso, Traylor, Trinastic, Tringali,
  Trozzo, Tsang, Tse, Tso, Tsukada, Tsuna, Tuyenbayev, Ueno, Ugolini,
  Unnikrishnan, Urban, Usman, Vahlbruch, Vajente, Valdes, Vallisneri, van
  Bakel, van Beuzekom, van~den Brand, Van Den~Broeck, Vander-Hyde, van~der
  Schaaf, van Heijningen, van Veggel, Vardaro, Varma, Vass, Vas\'uth, Vecchio,
  Vedovato, Veitch, Veitch, Venkateswara, Venugopalan, Verkindt, Vetrano,
  Vicer\'e, Viets, Vinciguerra, Vine, Vinet, Vitale, Vo, Vocca, Vorvick,
  Vyatchanin, Wade, Wade, Wade, Walet, Walker, Wallace, Walsh, Wang, Wang,
  Wang, Wang, Wang, Ward, Warner, Was, Watchi, Weaver, Wei, Weinert, Weinstein,
  Weiss, Wen, Wessel, We\ss{}els, Westerweck, Westphal, Wette, Whelan,
  Whitcomb, Whiting, Whittle, Wilken, Williams, Williams, Williamson, Willis,
  Willke, Wimmer, Winkler, Wipf, Wittel, Woan, Woehler, Wofford, Wong, Worden,
  Wright, Wu, Wysocki, Xiao, Yamamoto, Yancey, Yang, Yap, Yazback, Yu, Yu,
  Yvert, Zadro\ifmmode~\dot{z}\else \.{z}\fi{}ny, Zanolin, Zelenova, Zendri,
  Zevin, Zhang, Zhang, Zhang, Zhang, Zhao, Zhou, Zhou, Zhu, Zhu, Zimmerman,
  Zucker, \& Zweizig}]{PhysRevLett.119.161101}
Abbott, B.~P., Abbott, R., Abbott, T.~D., {et~al.} 2017{\natexlab{a}}, Phys.
  Rev. Lett., 119, 161101, \dodoi{10.1103/PhysRevLett.119.161101}

\bibitem[{Abbott {et~al.}(2017{\natexlab{b}})}]{LIGOScientific:2017ync}
Abbott, B.~P., {et~al.} 2017{\natexlab{b}}, Astrophys. J. Lett., 848, L12,
  \dodoi{10.3847/2041-8213/aa91c9}

\bibitem[{Abbott {et~al.}(2017{\natexlab{c}})}]{LIGOScientific:2017zic}
---. 2017{\natexlab{c}}, Astrophys. J. Lett., 848, L13,
  \dodoi{10.3847/2041-8213/aa920c}

\bibitem[{Abbott {et~al.}(2017{\natexlab{d}})}]{LIGOScientific:2017adf}
---. 2017{\natexlab{d}}, Nature, 551, 85, \dodoi{10.1038/nature24471}

\bibitem[{Abbott {et~al.}(2018{\natexlab{a}})}]{LIGOScientific:2018cki}
---. 2018{\natexlab{a}}, Phys. Rev. Lett., 121, 161101,
  \dodoi{10.1103/PhysRevLett.121.161101}

\bibitem[{Abbott {et~al.}(2018{\natexlab{b}})}]{KAGRA:2013rdx}
---. 2018{\natexlab{b}}, Living Rev. Rel., 21, 3,
  \dodoi{10.1007/s41114-020-00026-9}

\bibitem[{{Abbott} {et~al.}(2018){Abbott}, {Abbott}, {Abbott}, {Abernathy},
  {Acernese}, {Ackley}, {Adams}, {Adams}, {Addesso}, {Adhikari}, {Adya},
  {Affeldt}, {Agathos}, {Agatsuma}, {Aggarwal}, {Aguiar}, {Aiello}, {Ain},
  {Ajith}, {Akutsu}, {Allen}, {Allocca}, {Altin}, {Ananyeva}, {Anderson},
  {Anderson}, {Ando}, {Appert}, {Arai}, {Araya}, {Araya}, {Areeda}, {Arnaud},
  {Arun}, {Asada}, {Ascenzi}, {Ashton}, {Aso}, {Ast}, {Aston}, {Astone},
  {Atsuta}, {Aufmuth}, {Aulbert}, {Avila-Alvarez}, {Awai}, {Babak}, {Bacon},
  {Bader}, {Baiotti}, {Baker}, {Baldaccini}, {Ballardin}, {Ballmer},
  {Barayoga}, {Barclay}, {Barish}, {Barker}, {Barone}, {Barr}, {Barsotti},
  {Barsuglia}, {Barta}, {Bartlett}, {Barton}, {Bartos}, {Bassiri}, {Basti},
  {Batch}, {Baune}, {Bavigadda}, {Bazzan}, {B{\'e}csy}, {Beer}, {Bejger},
  {Belahcene}, {Belgin}, {Bell}, {Berger}, {Bergmann}, {Berry}, {Bersanetti},
  {Bertolini}, {Betzwieser}, {Bhagwat}, {Bhandare}, {Bilenko}, {Billingsley},
  {Billman}, {Birch}, {Birney}, {Birnholtz}, {Biscans}, {Bisht}, {Bitossi},
  {Biwer}, {Bizouard}, {Blackburn}, {Blackman}, {Blair}, {Blair}, {Blair},
  {Bloemen}, {Bock}, {Boer}, {Bogaert}, {Bohe}, {Bondu}, {Bonnand}, {Boom},
  {Bork}, {Boschi}, {Bose}, {Bouffanais}, {Bozzi}, {Bradaschia}, {Brady},
  {Braginsky}, {Branchesi}, {Brau}, {Briant}, {Brillet}, {Brinkmann},
  {Brisson}, {Brockill}, {Broida}, {Brooks}, {Brown}, {Brown}, {Brown},
  {Brunett}, {Buchanan}, {Buikema}, {Bulik}, {Bulten}, {Buonanno}, {Buskulic},
  {Buy}, {Byer}, {Cabero}, {Cadonati}, {Cagnoli}, {Cahillane}, {Calder{\'o}n
  Bustillo}, {Callister}, {Calloni}, {Camp}, {Cannon}, {Cao}, {Cao}, {Capano},
  {Capocasa}, {Carbognani}, {Caride}, {Casanueva Diaz}, {Casentini}, {Caudill},
  {Cavagli{\`a}}, {Cavalier}, {Cavalieri}, {Cella}, {Cepeda}, {Cerboni
  Baiardi}, {Cerretani}, {Cesarini}, {Chamberlin}, {Chan}, {Chao}, {Charlton},
  {Chassande-Mottin}, {Cheeseboro}, {Chen}, {Chen}, {Cheng}, {Chincarini},
  {Chiummo}, {Chmiel}, {Cho}, {Cho}, {Chow}, {Christensen}, {Chu}, {Chua},
  {Chua}, {Chung}, {Ciani}, {Clara}, {Clark}, {Cleva}, {Cocchieri}, {Coccia},
  {Cohadon}, {Colla}, {Collette}, {Cominsky}, {Constancio}, {Conti}, {Cooper},
  {Corbitt}, {Cornish}, {Corsi}, {Cortese}, {Costa}, {Coughlin}, {Coughlin},
  {Coulon}, {Countryman}, {Couvares}, {Covas}, {Cowan}, {Coward}, {Cowart},
  {Coyne}, {Coyne}, {Creighton}, {Creighton}, {Cripe}, {Crowder}, {Cullen},
  {Cumming}, {Cunningham}, {Cuoco}, {Dal Canton}, {Danilishin}, {D'Antonio},
  {Danzmann}, {Dasgupta}, {da Silva Costa}, {Dattilo}, {Dave}, {Davier},
  {Davies}, {Davis}, {Daw}, {Day}, {Day}, {de}, {Debra}, {Debreczeni},
  {Degallaix}, {de Laurentis}, {Del{\'e}glise}, {Del Pozzo}, {Denker}, {Dent},
  {Dergachev}, {De Rosa}, {Derosa}, {Desalvo}, {Devine}, {Dhurandhar},
  {D{\'\i}az}, {di Fiore}, {di Giovanni}, {di Girolamo}, {di Lieto}, {di Pace},
  {di Palma}, {di Virgilio}, {Doctor}, {Doi}, {Dolique}, {Donovan}, {Dooley},
  {Doravari}, {Dorrington}, {Douglas}, {Dovale {\'A}lvarez}, {Downes}, {Drago},
  {Drever}, {Driggers}, {Du}, {Ducrot}, {Dwyer}, {Eda}, {Edo}, {Edwards},
  {Effler}, {Eggenstein}, {Ehrens}, {Eichholz}, {Eikenberry}, {Eisenstein},
  {Essick}, {Etienne}, {Etzel}, {Evans}, {Evans}, {Everett}, {Factourovich},
  {Fafone}, {Fair}, {Fairhurst}, {Fan}, {Farinon}, {Farr}, {Farr},
  {Fauchon-Jones}, {Favata}, {Fays}, {Fehrmann}, {Fejer}, {Fern{\'a}ndez
  Galiana}, {Ferrante}, {Ferreira}, {Ferrini}, {Fidecaro}, {Fiori}, {Fiorucci},
  {Fisher}, {Flaminio}, {Fletcher}, {Fong}, {Forsyth}, {Fournier}, {Frasca},
  {Frasconi}, {Frei}, {Freise}, {Frey}, {Frey}, {Fries}, {Fritschel}, {Frolov},
  {Fujii}, {Fujimoto}, {Fulda}, {Fyffe}, {Gabbard}, {Gadre}, {Gaebel}, {Gair},
  {Gammaitoni}, {Gaonkar}, {Garufi}, {Gaur}, {Gayathri}, {Gehrels}, {Gemme},
  {Genin}, {Gennai}, {George}, {Gergely}, {Germain}, {Ghonge}, {Ghosh},
  {Ghosh}, {Ghosh}, {Giaime}, {Giardina}, {Giazotto}, {Gill}, {Glaefke},
  {Goetz}, {Goetz}, {Gondan}, {Gonz{\'a}lez}, {Gonzalez Castro}, {Gopakumar},
  {Gorodetsky}, {Gossan}, {Gosselin}, {Gouaty}, {Grado}, {Graef}, {Granata},
  {Grant}, {Gras}, {Gray}, {Greco}, {Green}, {Groot}, {Grote}, {Grunewald},
  {Guidi}, {Guo}, {Gupta}, {Gupta}, {Gushwa}, {Gustafson}, {Gustafson},
  {Hacker}, {Hagiwara}, {Hall}, {Hall}, {Hammond}, {Haney}, {Hanke}, {Hanks},
  {Hanna}, {Hannam}, {Hanson}, {Hardwick}, {Harms}, {Harry}, {Harry}, {Hart},
  {Hartman}, {Haster}, {Haughian}, {Hayama}, {Healy}, {Heidmann}, {Heintze},
  {Heitmann}, {Hello}, {Hemming}, {Hendry}, {Heng}, {Hennig}, {Henry},
  {Heptonstall}, {Heurs}, {Hild}, {Hirose}, {Hoak}, {Hofman}, {Holt}, {Holz},
  {Hopkins}, {Hough}, {Houston}, {Howell}, {Hu}, {Huerta}, {Huet}, {Hughey},
  {Husa}, {Huttner}, {Huynh-Dinh}, {Indik}, {Ingram}, {Inta}, {Ioka}, {Isa},
  {Isac}, {Isi}, {Isogai}, {Itoh}, {Iyer}, {Izumi}, {Jacqmin}, {Jani},
  {Jaranowski}, {Jawahar}, {Jim{\'e}nez-Forteza}, {Johnson}, {Jones}, {Jones},
  {Jonker}, {Ju}, {Junker}, {Kagawa}, {Kajita}, {Kakizaki}, {Kalaghatgi},
  {Kalogera}, {Kamiizumi}, {Kanda}, {Kandhasamy}, {Kanemura}, {Kaneyama},
  {Kang}, {Kanner}, {Karki}, {Karvinen}, {Kasprzack}, {Kataoka},
  {Katsavounidis}, {Katzman}, {Kaufer}, {Kaur}, {Kawabe}, {Kawai}, {Kawamura},
  {K{\'e}f{\'e}lian}, {Keitel}, {Kelley}, {Kennedy}, {Key}, {Khalili}, {Khan},
  {Khan}, {Khan}, {Khazanov}, {Kijbunchoo}, {Kim}, {Kim}, {Kim}, {Kim}, {Kim},
  {Kim}, {Kimbrell}, {Kimura}, {King}, {King}, {Kirchhoff}, {Kissel}, {Klein},
  {Kleybolte}, {Klimenko}, {Koch}, {Koehlenbeck}, {Kojima}, {Kokeyama},
  {Koley}, {Komori}, {Kondrashov}, {Kontos}, {Korobko}, {Korth}, {Kotake},
  {Kowalska}, {Kozak}, {Kr{\"a}mer}, {Kringel}, {Krishnan}, {Kr{\'o}lak},
  {Kuehn}, {Kumar}, {Kumar}, {Kumar}, {Kuo}, {Kuroda}, {Kutynia}, {Kuwahara},
  {Lackey}, {Landry}, {Lang}, {Lange}, {Lantz}, {Lanza}, {Lartaux-Vollard},
  {Lasky}, {Laxen}, {Lazzarini}, {Lazzaro}, {Leaci}, {Leavey}, {Lebigot},
  {Lee}, {Lee}, {Lee}, {Lee}, {Lee}, {Lehmann}, {Lenon}, {Leonardi}, {Leong},
  {Leroy}, {Letendre}, {Levin}, {Li}, {Libson}, {Littenberg}, {Liu},
  {Lockerbie}, {Lombardi}, {London}, {Lord}, {Lorenzini}, {Loriette},
  {Lormand}, {Losurdo}, {Lough}, {Lousto}, {Lovelace}, {L{\"u}ck}, {Lundgren},
  {Lynch}, {Ma}, {Macfoy}, {Machenschalk}, {Macinnis}, {MacLeod},
  {Maga{\~n}a-Sandoval}, {Majorana}, {Maksimovic}, {Malvezzi}, {Man}, {Mandic},
  {Mangano}, {Mano}, {Mansell}, {Manske}, {Mantovani}, {Marchesoni}, {Marchio},
  {Marion}, {M{\'a}rka}, {M{\'a}rka}, {Markosyan}, {Maros}, {Martelli},
  {Martellini}, {Martin}, {Martynov}, {Mason}, {Masserot}, {Massinger},
  {Masso-Reid}, {Mastrogiovanni}, {Matichard}, {Matone}, {Matsumoto},
  {Matsushima}, {Mavalvala}, {Mazumder}, {McCarthy}, {McClelland}, {McCormick},
  {McGrath}, {McGuire}, {McIntyre}, {McIver}, {McManus}, {McRae}, {McWilliams},
  {Meacher}, {Meadors}, {Meidam}, {Melatos}, {Mendell}, {Mendoza-Gandara},
  {Mercer}, {Merilh}, {Merzougui}, {Meshkov}, {Messenger}, {Messick},
  {Metzdorff}, {Meyers}, {Mezzani}, {Miao}, {Michel}, {Michimura}, {Middleton},
  {Mikhailov}, {Milano}, {Miller}, {Miller}, {Miller}, {Miller}, {Millhouse},
  {Minenkov}, {Ming}, {Mirshekari}, {Mishra}, {Mitrofanov}, {Mitselmakher},
  {Mittleman}, {Miyakawa}, {Miyamoto}, {Miyamoto}, {Miyoki}, {Moggi}, {Mohan},
  {Mohapatra}, {Montani}, {Moore}, {Moore}, {Moraru}, {Moreno}, {Morii},
  {Morisaki}, {Moriwaki}, {Morriss}, {Mours}, {Mow-Lowry}, {Mueller}, {Muir},
  {Mukherjee}, {Mukherjee}, {Mukherjee}, {Mukund}, {Mullavey}, {Munch},
  {Muniz}, {Murray}, {Mytidis}, {Nagano}, {Nakamura}, {Nakamura}, {Nakano},
  {Nakano}, {Nakano}, {Nakao}, {Napier}, {Nardecchia}, {Narikawa},
  {Naticchioni}, {Nelemans}, {Nelson}, {Neri}, {Nery}, {Neunzert}, {Newport},
  {Newton}, {Nguyen}, {Ni}, {Nielsen}, {Nissanke}, {Nitz}, {Noack}, {Nocera},
  {Nolting}, {Normandin}, {Nuttall}, {Oberling}, {Ochsner}, {Oelker}, {Ogin},
  {Oh}, {Oh}, {Ohashi}, {Ohishi}, {Ohkawa}, {Ohme}, {Okutomi}, {Oliver}, {Ono},
  {Ono}, {Oohara}, {Oppermann}, {Oram}, {O'Reilly}, {O'Shaughnessy}, {Ottaway},
  {Overmier}, {Owen}, {Pace}, {Page}, {Pai}, {Pai}, {Palamos}, {Palashov},
  {Palomba}, {Pal-Singh}, {Pan}, {Pankow}, {Pannarale}, {Pant}, {Paoletti},
  {Paoli}, {Papa}, {Paris}, {Parker}, {Pascucci}, {Pasqualetti}, {Passaquieti},
  {Passuello}, {Patricelli}, {Pearlstone}, {Pedraza}, {Pedurand}, {Pekowsky},
  {Pele}, {Pe{\~n}a Arellano}, {Penn}, {Perez}, {Perreca}, {Perri}, {Pfeiffer},
  {Phelps}, {Piccinni}, {Pichot}, {Piergiovanni}, {Pierro}, {Pillant},
  {Pinard}, {Pinto}, {Pitkin}, {Poe}, {Poggiani}, {Popolizio}, {Post},
  {Powell}, {Prasad}, {Pratt}, {Predoi}, {Prestegard}, {Prijatelj}, {Principe},
  {Privitera}, {Prodi}, {Prokhorov}, {Puncken}, {Punturo}, {Puppo},
  {P{\"u}rrer}, {Qi}, {Qin}, {Qiu}, {Quetschke}, {Quintero}, {Quitzow-James},
  {Raab}, {Rabeling}, {Radkins}, {Raffai}, {Raja}, {Rajan}, {Rakhmanov},
  {Rapagnani}, {Raymond}, {Razzano}, {Re}, {Read}, {Regimbau}, {Rei}, {Reid},
  {Reitze}, {Rew}, {Reyes}, {Rhoades}, {Ricci}, {Riles}, {Rizzo}, {Robertson},
  {Robie}, {Robinet}, {Rocchi}, {Rolland}, {Rollins}, {Roma}, {Romano},
  {Romie}, {Rosi{\'n}ska}, {Rowan}, {R{\"u}diger}, {Ruggi}, {Ryan}, {Sachdev},
  {Sadecki}, {Sadeghian}, {Sago}, {Saijo}, {Saito}, {Sakai}, {Sakellariadou},
  {Salconi}, {Saleem}, {Salemi}, {Samajdar}, {Sammut}, {Sampson}, {Sanchez},
  {Sandberg}, {Sanders}, {Sasaki}, {Sassolas}, {Sathyaprakash}, {Sato}, {Sato},
  {Saulson}, {Sauter}, {Savage}, {Sawadsky}, {Schale}, {Scheuer}, {Schmidt},
  {Schmidt}, {Schmidt}, {Schnabel}, {Schofield}, {Sch{\"o}nbeck}, {Schreiber},
  {Schuette}, {Schutz}, {Schwalbe}, {Scott}, {Scott}, {Sekiguchi}, {Sekiguchi},
  {Sellers}, {Sengupta}, {Sentenac}, {Sequino}, {Sergeev}, {Setyawati},
  {Shaddock}, {Shaffer}, {Shahriar}, {Shapiro}, {Shawhan}, {Sheperd},
  {Shibata}, {Shikano}, {Shimoda}, {Shoda}, {Shoemaker}, {Shoemaker},
  {Siellez}, {Siemens}, {Sieniawska}, {Sigg}, {Silva}, {Singer}, {Singer},
  {Singh}, {Singh}, {Singhal}, {Sintes}, {Slagmolen}, {Smith}, {Smith},
  {Smith}, {Somiya}, {Son}, {Sorazu}, {Sorrentino}, {Souradeep}, {Spencer},
  {Srivastava}, {Staley}, {Steinke}, {Steinlechner}, {Steinlechner},
  {Steinmeyer}, {Stephens}, {Stevenson}, {Stone}, {Strain}, {Straniero},
  {Stratta}, {Strigin}, {Sturani}, {Stuver}, {Sugimoto}, {Summerscales}, {Sun},
  {Sunil}, {Sutton}, {Suzuki}, {Swinkels}, {Szczepa{\'n}czyk}, {Tacca},
  {Tagoshi}, {Takada}, {Takahashi}, {Takahashi}, {Takamori}, {Talukder},
  {Tanaka}, {Tanaka}, {Tanaka}, {Tanner}, {T{\'a}pai}, {Taracchini}, {Tatsumi},
  {Taylor}, {Telada}, {Theeg}, {Thomas}, {Thomas}, {Thomas}, {Thorne},
  {Thrane}, {Tippens}, {Tiwari}, {Tiwari}, {Tokmakov}, {Toland}, {Tomaru},
  {Tomlinson}, {Tonelli}, {Tornasi}, {Torrie}, {T{\"o}yr{\"a}}, {Travasso},
  {Traylor}, {Trifir{\`o}}, {Trinastic}, {Tringali}, {Trozzo}, {Tse}, {Tso},
  {Tsubono}, {Tsuzuki}, {Turconi}, {Tuyenbayev}, {Uchiyama}, {Uehara}, {Ueki},
  {Ueno}, {Ugolini}, {Unnikrishnan}, {Urban}, {Ushiba}, {Usman}, {Vahlbruch},
  {Vajente}, {Valdes}, {van Bakel}, {van Beuzekom}, {van den Brand}, {van den
  Broeck}, {Vander-Hyde}, {van der Schaaf}, {van Heijningen}, {van Putten},
  {van Veggel}, {Vardaro}, {Varma}, {Vass}, {Vas{\'u}th}, {Vecchio},
  {Vedovato}, {Veitch}, {Veitch}, {Venkateswara}, {Venugopalan}, {Verkindt},
  {Vetrano}, {Vicer{\'e}}, {Viets}, {Vinciguerra}, {Vine}, {Vinet}, {Vitale},
  {Vo}, {Vocca}, {Vorvick}, {Voss}, {Vousden}, {Vyatchanin}, {Wade}, {Wade},
  {Wade}, {Wakamatsu}, {Walker}, {Wallace}, {Walsh}, {Wang}, {Wang}, {Wang},
  {Wang}, {Ward}, {Warner}, {Was}, {Watchi}, {Weaver}, {Wei}, {Weinert},
  {Weinstein}, {Weiss}, {Wen}, {We{\ss}els}, {Westphal}, {Wette}, {Whelan},
  {Whiting}, {Whittle}, {Williams}, {Williams}, {Williamson}, {Willis},
  {Willke}, {Wimmer}, {Winkler}, {Wipf}, {Wittel}, {Woan}, {Woehler}, {Worden},
  {Wright}, {Wu}, {Wu}, {Yam}, {Yamamoto}, {Yamamoto}, {Yamamoto}, {Yancey},
  {Yano}, {Yap}, {Yokoyama}, {Yokozawa}, {Yoon}, {Yu}, {Yu}, {Yuzurihara},
  {Yvert}, {Zadro{\.z}ny}, {Zangrando}, {Zanolin}, {Zeidler}, {Zendri},
  {Zevin}, {Zhang}, {Zhang}, {Zhang}, {Zhang}, {Zhao}, {Zhou}, {Zhou}, {Zhu},
  {Zhu}, {Zucker}, {Zweizig}, {Kagra Collaboration}, \& {VIRGO
  Collaboration}}]{2018LRR....21....3A}
{Abbott}, B.~P., {Abbott}, R., {Abbott}, T.~D., {et~al.} 2018, Living Reviews
  in Relativity, 21, 3, \dodoi{10.1007/s41114-018-0012-9}

\bibitem[{Abbott {et~al.}(2021{\natexlab{a}})}]{LIGOScientific:2019zcs}
Abbott, B.~P., {et~al.} 2021{\natexlab{a}}, Astrophys. J., 909, 218,
  \dodoi{10.3847/1538-4357/abdcb7}

\bibitem[{Abbott {et~al.}(2021{\natexlab{b}})}]{LIGOScientific:2021djp}
Abbott, R., {et~al.} 2021{\natexlab{b}}.
\newblock \doarXiv{2111.03606}

\bibitem[{Abbott {et~al.}(2021{\natexlab{c}})}]{LIGOScientific:2021psn}
---. 2021{\natexlab{c}}.
\newblock \doarXiv{2111.03634}

\bibitem[{Abbott {et~al.}(2021{\natexlab{d}})}]{LIGOScientific:2021sio}
---. 2021{\natexlab{d}}.
\newblock \doarXiv{2112.06861}

\bibitem[{Abbott {et~al.}(2021{\natexlab{e}})}]{LIGOScientific:2021aug}
---. 2021{\natexlab{e}}.
\newblock \doarXiv{2111.03604}

\bibitem[{{Abbott} {et~al.}(2021){Abbott}, {Abbott}, {Abraham}, {Acernese},
  {Ackley}, {Adams}, {Adams}, {Adhikari}, {Adya}, {Affeldt}, {Agathos},
  {Agatsuma}, {Aggarwal}, {Aguiar}, {Aiello}, {Ain}, {Ajith}, {Allen},
  {Allocca}, {Altin}, {Amato}, {Anand}, {Ananyeva}, {Anderson}, {Anderson},
  {Angelova}, {Ansoldi}, {Antelis}, {Antier}, {Appert}, {Arai}, {Araya},
  {Areeda}, {Ar{\`e}ne}, {Arnaud}, {Aronson}, {Arun}, {Asali}, {Ascenzi},
  {Ashton}, {Aston}, {Astone}, {Aubin}, {Aufmuth}, {AultONeal}, {Austin},
  {Avendano}, {Babak}, {Badaracco}, {Bader}, {Bae}, {Baer}, {Bagnasco},
  {Baird}, {Ball}, {Ballardin}, {Ballmer}, {Bals}, {Balsamo}, {Baltus},
  {Banagiri}, {Bankar}, {Bankar}, {Barayoga}, {Barbieri}, {Barish}, {Barker},
  {Barneo}, {Barnum}, {Barone}, {Barr}, {Barsotti}, {Barsuglia}, {Barta},
  {Bartlett}, {Bartos}, {Bassiri}, {Basti}, {Bawaj}, {Bayley}, {Bazzan},
  {Becher}, {B{\'e}csy}, {Bedakihale}, {Bejger}, {Belahcene}, {Beniwal},
  {Benjamin}, {Bennett}, {Bentley}, {Bergamin}, {Berger}, {Bergmann},
  {Bernuzzi}, {Berry}, {Bersanetti}, {Bertolini}, {Betzwieser}, {Bhandare},
  {Bhandari}, {Bhattacharjee}, {Bidler}, {Bilenko}, {Billingsley}, {Birney},
  {Birnholtz}, {Biscans}, {Bischi}, {Biscoveanu}, {Bisht}, {Bitossi},
  {Bizouard}, {Blackburn}, {Blackman}, {Blair}, {Blair}, {Blair}, {Blanch},
  {Bobba}, {Bode}, {Boer}, {Boetzel}, {Bogaert}, {Boldrini}, {Bondu},
  {Bonilla}, {Bonnand}, {Booker}, {Boom}, {Bork}, {Boschi}, {Bose},
  {Bossilkov}, {Boudart}, {Bouffanais}, {Bozzi}, {Bradaschia}, {Brady},
  {Bramley}, {Branchesi}, {Brau}, {Breschi}, {Briant}, {Briggs}, {Brighenti},
  {Brillet}, {Brinkmann}, {Brockill}, {Brooks}, {Brooks}, {Brown}, {Brunett},
  {Bruno}, {Bruntz}, {Buikema}, {Bulik}, {Bulten}, {Buonanno}, {Buscicchio},
  {Buskulic}, {Byer}, {Cabero}, {Cadonati}, {Caesar}, {Cagnoli}, {Cahillane},
  {Calder{\'o}n Bustillo}, {Callaghan}, {Callister}, {Calloni}, {Camp},
  {Canepa}, {Cannon}, {Cao}, {Cao}, {Carapella}, {Carbognani}, {Carney},
  {Carpinelli}, {Carullo}, {Carver}, {Casanueva Diaz}, {Casentini}, {Caudill},
  {Cavagli{\`a}}, {Cavalier}, {Cavalieri}, {Cella}, {Cerd{\'a}-Dur{\'a}n},
  {Cesarini}, {Chaibi}, {Chakravarti}, {Chan}, {Chan}, {Chandra}, {Chanial},
  {Chao}, {Charlton}, {Chase}, {Chassande-Mottin}, {Chatterjee},
  {Chattopadhyay}, {Chaturvedi}, {Chatziioannou}, {Chen}, {Chen}, {Chen},
  {Chen}, {Cheng}, {Cheong}, {Chia}, {Chiadini}, {Chierici}, {Chincarini},
  {Chiummo}, {Cho}, {Cho}, {Cho}, {Choate}, {Christensen}, {Chu}, {Chua},
  {Chung}, {Chung}, {Ciani}, {Ciecielag}, {Cie{\'s}lar}, {Cifaldi}, {Ciobanu},
  {Ciolfi}, {Cipriano}, {Cirone}, {Clara}, {Clark}, {Clark}, {Clarke},
  {Clearwater}, {Clesse}, {Cleva}, {Coccia}, {Cohadon}, {Cohen}, {Colleoni},
  {Collette}, {Collins}, {Colpi}, {Constancio}, {Conti}, {Cooper}, {Corban},
  {Corbitt}, {Cordero-Carri{\'o}n}, {Corezzi}, {Corley}, {Cornish}, {Corre},
  {Corsi}, {Cortese}, {Costa}, {Cotesta}, {Coughlin}, {Coughlin}, {Coulon},
  {Countryman}, {Couvares}, {Covas}, {Coward}, {Cowart}, {Coyne}, {Coyne},
  {Creighton}, {Creighton}, {Croquette}, {Crowder}, {Cudell}, {Cullen},
  {Cumming}, {Cummings}, {Cunningham}, {Cuoco}, {Curylo}, {Dal Canton},
  {D{\'a}lya}, {Dana}, {DaneshgaranBajastani}, {D'Angelo}, {Danilishin},
  {D'Antonio}, {Danzmann}, {Darsow-Fromm}, {Dasgupta}, {Datrier}, {Dattilo},
  {Dave}, {Davier}, {Davies}, {Davis}, {Daw}, {Dean}, {DeBra}, {Deenadayalan},
  {Degallaix}, {De Laurentis}, {Del{\'e}glise}, {Del Favero}, {De Lillo}, {De
  Lillo}, {Del Pozzo}, {DeMarchi}, {De Matteis}, {D'Emilio}, {Demos}, {Denker},
  {Dent}, {Depasse}, {De Pietri}, {De Rosa}, {De Rossi}, {DeSalvo}, {de
  Varona}, {Dhurandhar}, {D{\'\i}az}, {Diaz-Ortiz}, {Didio}, {Dietrich}, {Di
  Fiore}, {DiFronzo}, {Di Giorgio}, {Di Giovanni}, {Di Giovanni}, {Di
  Girolamo}, {Di Lieto}, {Ding}, {Di Pace}, {Di Palma}, {Di Renzo},
  {Divakarla}, {Dmitriev}, {Doctor}, {D'Onofrio}, {Donovan}, {Dooley},
  {Doravari}, {Dorrington}, {Downes}, {Drago}, {Driggers}, {Du}, {Ducoin},
  {Dupej}, {Durante}, {D'Urso}, {Duverne}, {Dwyer}, {Easter}, {Eddolls},
  {Edelman}, {Edo}, {Edy}, {Effler}, {Eichholz}, {Eikenberry}, {Eisenmann},
  {Eisenstein}, {Ejlli}, {Errico}, {Essick}, {Estell{\'e}s}, {Estevez},
  {Etienne}, {Etzel}, {Evans}, {Evans}, {Ewing}, {Fafone}, {Fair}, {Fairhurst},
  {Fan}, {Farah}, {Farinon}, {Farr}, {Farr}, {Fauchon-Jones}, {Favata}, {Fays},
  {Fazio}, {Feicht}, {Fejer}, {Feng}, {Fenyvesi}, {Ferguson},
  {Fernandez-Galiana}, {Ferrante}, {Ferreira}, {Fidecaro}, {Figura}, {Fiori},
  {Fiorucci}, {Fishbach}, {Fisher}, {Fishner}, {Fittipaldi}, {Fitz-Axen},
  {Fiumara}, {Flaminio}, {Floden}, {Flynn}, {Fong}, {Font}, {Forsyth},
  {Fournier}, {Frasca}, {Frasconi}, {Frei}, {Freise}, {Frey}, {Frey},
  {Fritschel}, {Frolov}, {Fronz{\'e}}, {Fulda}, {Fyffe}, {Gabbard}, {Gadre},
  {Gaebel}, {Gair}, {Gais}, {Galaudage}, {Gamba}, {Ganapathy}, {Ganguly},
  {Gaonkar}, {Garaventa}, {Garc{\'\i}a-Quir{\'o}s}, {Garufi}, {Gateley},
  {Gaudio}, {Gayathri}, {Gemme}, {Gennai}, {George}, {George}, {Gergely},
  {Ghonge}, {Ghosh}, {Ghosh}, {Ghosh}, {Giacomazzo}, {Giacoppo}, {Giaime},
  {Giardina}, {Gibson}, {Gier}, {Gill}, {Giri}, {Glanzer}, {Gleckl}, {Godwin},
  {Goetz}, {Goetz}, {Gohlke}, {Goncharov}, {Gonz{\'a}lez}, {Gopakumar},
  {Gossan}, {Gosselin}, {Gouaty}, {Grace}, {Grado}, {Granata}, {Granata},
  {Grant}, {Gras}, {Grassia}, {Gray}, {Gray}, {Greco}, {Green}, {Green},
  {Gretarsson}, {Griggs}, {Grignani}, {Grimaldi}, {Grimes}, {Grimm}, {Grote},
  {Grunewald}, {Gruning}, {Guerrero}, {Guidi}, {Guimaraes}, {Guix{\'e}},
  {Gulati}, {Guo}, {Gupta}, {Gupta}, {Gupta}, {Gustafson}, {Gustafson},
  {Guzman}, {Haegel}, {Halim}, {Hall}, {Hamilton}, {Hammond}, {Haney}, {Hanke},
  {Hanks}, {Hanna}, {Hannuksela}, {Hannuksela}, {Hansen}, {Hansen}, {Hanson},
  {Harder}, {Hardwick}, {Haris}, {Harms}, {Harry}, {Harry}, {Hartwig},
  {Hasskew}, {Haster}, {Haughian}, {Hayes}, {Healy}, {Heidmann}, {Heintze},
  {Heinze}, {Heinzel}, {Heitmann}, {Hellman}, {Hello}, {Helmling-Cornell},
  {Hemming}, {Hendry}, {Heng}, {Hennes}, {Hennig}, {Hennig}, {Hernandez
  Vivanco}, {Heurs}, {Hild}, {Hill}, {Hines}, {Hochheim}, {Hofgard}, {Hofman},
  {Hohmann}, {Holgado}, {Holland}, {Hollows}, {Holmes}, {Holt}, {Holz},
  {Hopkins}, {Horst}, {Hough}, {Howell}, {Hoy}, {Hoyland}, {Huang},
  {H{\"u}bner}, {Huddart}, {Huerta}, {Hughey}, {Hui}, {Husa}, {Huttner},
  {Hutzler}, {Huxford}, {Huynh-Dinh}, {Idzkowski}, {Iess}, {Imperato},
  {Inchauspe}, {Ingram}, {Intini}, {Isi}, {Iyer}, {JaberianHamedan}, {Jacqmin},
  {Jadhav}, {Jadhav}, {James}, {Jani}, {Janssens}, {Janthalur}, {Jaranowski},
  {Jariwala}, {Jaume}, {Jenkins}, {Jeunon}, {Jiang}, {Johns}, {Jones}, {Jones},
  {Jones}, {Jones}, {Jones}, {Jonker}, {Ju}, {Junker}, {Kalaghatgi},
  {Kalogera}, {Kamai}, {Kandhasamy}, {Kang}, {Kanner}, {Kapadia}, {Kapasi},
  {Karathanasis}, {Karki}, {Kashyap}, {Kasprzack}, {Kastaun}, {Katsanevas},
  {Katsavounidis}, {Katzman}, {Kawabe}, {K{\'e}f{\'e}lian}, {Keitel}, {Key},
  {Khadka}, {Khalili}, {Khan}, {Khan}, {Khazanov}, {Khetan}, {Khursheed},
  {Kijbunchoo}, {Kim}, {Kim}, {Kim}, {Kim}, {Kim}, {Kim}, {Kimball}, {King},
  {Kinley-Hanlon}, {Kirchhoff}, {Kissel}, {Kleybolte}, {Klimenko}, {Knowles},
  {Knyazev}, {Koch}, {Koehlenbeck}, {Koekoek}, {Koley}, {Kolstein}, {Komori},
  {Kondrashov}, {Kontos}, {Koper}, {Korobko}, {Korth}, {Kovalam}, {Kozak},
  {Kr{\"a}mer}, {Kringel}, {Krishnendu}, {Kr{\'o}lak}, {Kuehn}, {Kumar},
  {Kumar}, {Kumar}, {Kumar}, {Kuns}, {Kwang}, {Lackey}, {Laghi}, {Lalande},
  {Lam}, {Lamberts}, {Landry}, {Lane}, {Lang}, {Lange}, {Lantz}, {Lanza}, {La
  Rosa}, {Lartaux-Vollard}, {Lasky}, {Laxen}, {Lazzarini}, {Lazzaro}, {Leaci},
  {Leavey}, {Lecoeuche}, {Lee}, {Lee}, {Lee}, {Lee}, {Lehmann}, {Leon},
  {Leroy}, {Letendre}, {Levin}, {Li}, {Li}, {Li}, {Li}, {Li}, {Linde},
  {Linker}, {Linley}, {Littenberg}, {Liu}, {Liu}, {Llorens-Monteagudo}, {Lo},
  {Lockwood}, {London}, {Longo}, {Lorenzini}, {Loriette}, {Lormand}, {Losurdo},
  {Lough}, {Lousto}, {Lovelace}, {L{\"u}ck}, {Lumaca}, {Lundgren}, {Ma},
  {Macas}, {MacInnis}, {Macleod}, {MacMillan}, {Macquet}, {Maga{\~n}a
  Hernandez}, {Maga{\~n}a-Sandoval}, {Magazz{\`u}}, {Magee}, {Majorana},
  {Maksimovic}, {Maliakal}, {Malik}, {Man}, {Mandic}, {Mangano}, {Mansell},
  {Manske}, {Mantovani}, {Mapelli}, {Marchesoni}, {Marion}, {M{\'a}rka},
  {M{\'a}rka}, {Markakis}, {Markosyan}, {Markowitz}, {Maros}, {Marquina},
  {Marsat}, {Martelli}, {Martin}, {Martin}, {Martinez}, {Martinez}, {Martynov},
  {Masalehdan}, {Mason}, {Massera}, {Masserot}, {Massinger}, {Masso-Reid},
  {Mastrogiovanni}, {Matas}, {Mateu-Lucena}, {Matichard}, {Matiushechkina},
  {Mavalvala}, {Maynard}, {McCann}, {McCarthy}, {McClelland}, {McCormick},
  {McCuller}, {McGuire}, {McIsaac}, {McIver}, {McManus}, {McRae}, {McWilliams},
  {Meacher}, {Meadors}, {Mehmet}, {Mehta}, {Melatos}, {Melchor}, {Mendell},
  {Menendez-Vazquez}, {Mercer}, {Mereni}, {Merfeld}, {Merilh}, {Merritt},
  {Merzougui}, {Meshkov}, {Messenger}, {Messick}, {Metzdorff}, {Meyers},
  {Meylahn}, {Mhaske}, {Miani}, {Miao}, {Michaloliakos}, {Michel}, {Middleton},
  {Milano}, {Miller}, {Miller}, {Millhouse}, {Mills}, {Milotti},
  {Milovich-Goff}, {Minazzoli}, {Minenkov}, {Mir}, {Mishkin}, {Mishra},
  {Mistry}, {Mitra}, {Mitrofanov}, {Mitselmakher}, {Mittleman}, {Mo},
  {Mogushi}, {Mohapatra}, {Mohite}, {Molina}, {Molina-Ruiz}, {Mondin},
  {Montani}, {Moore}, {Moraru}, {Morawski}, {Moreno}, {Morisaki}, {Mours},
  {Mow-Lowry}, {Mozzon}, {Muciaccia}, {Mukherjee}, {Mukherjee}, {Mukherjee},
  {Mukherjee}, {Mukund}, {Mullavey}, {Munch}, {Mu{\~n}iz}, {Murray}, {Nadji},
  {Nagar}, {Nardecchia}, {Naticchioni}, {Nayak}, {Neil}, {Neilson}, {Nelemans},
  {Nelson}, {Nery}, {Neunzert}, {Ng}, {Ng}, {Nguyen}, {Nguyen}, {Nguyen},
  {Nichols}, {Nissanke}, {Nocera}, {Noh}, {North}, {Nothard}, {Nuttall},
  {Oberling}, {O'Brien}, {O'Dell}, {Oganesyan}, {Ogin}, {Oh}, {Oh}, {Ohme},
  {Ohta}, {Okada}, {Olivetto}, {Oppermann}, {Oram}, {O'Reilly}, {Ormiston},
  {Ormsby}, {Ortega}, {O'Shaughnessy}, {Ossokine}, {Osthelder}, {Ottaway},
  {Overmier}, {Owen}, {Pace}, {Pagano}, {Page}, {Pagliaroli}, {Pai}, {Pai},
  {Palamos}, {Palashov}, {Palomba}, {Pan}, {Panda}, {Pang}, {Pankow},
  {Pannarale}, {Pant}, {Paoletti}, {Paoli}, {Paolone}, {Parker}, {Pascucci},
  {Pasqualetti}, {Passaquieti}, {Passuello}, {Patel}, {Patricelli}, {Payne},
  {Pechsiri}, {Pedraza}, {Pegoraro}, {Pele}, {Penn}, {Perego}, {Perez},
  {P{\'e}rigois}, {Perreca}, {Perri{\`e}s}, {Petermann}, {Petterson},
  {Pfeiffer}, {Pham}, {Phukon}, {Piccinni}, {Pichot}, {Piendibene},
  {Piergiovanni}, {Pierini}, {Pierro}, {Pillant}, {Pilo}, {Pinard}, {Pinto},
  {Piotrzkowski}, {Pirello}, {Pitkin}, {Placidi}, {Plastino}, {Pluchar},
  {Poggiani}, {Polini}, {Pong}, {Ponrathnam}, {Popolizio}, {Porter},
  {Poverman}, {Powell}, {Pracchia}, {Prajapati}, {Prasai}, {Prasanna},
  {Pratten}, {Prestegard}, {Principe}, {Prodi}, {Prokhorov}, {Prosposito},
  {Puecher}, {Punturo}, {Puosi}, {Puppo}, {P{\"u}rrer}, {Qi}, {Quetschke},
  {Quinonez}, {Quitzow-James}, {Raab}, {Raaijmakers}, {Radkins}, {Radulesco},
  {Raffai}, {Rafferty}, {Rail}, {Raja}, {Rajan}, {Rajbhandari}, {Rakhmanov},
  {Ramirez}, {Ramirez}, {Ramos-Buades}, {Rana}, {Rao}, {Rapagnani}, {Rapol},
  {Ratto}, {Raymond}, {Razzano}, {Read}, {Regimbau}, {Rei}, {Reid}, {Reitze},
  {Rettegno}, {Ricci}, {Richardson}, {Richardson}, {Richardson}, {Ricker},
  {Riemenschneider}, {Riles}, {Rizzo}, {Robertson}, {Robinet}, {Rocchi},
  {Rocha}, {Rodriguez}, {Rodriguez-Soto}, {Rolland}, {Rollins}, {Roma},
  {Romanelli}, {Romano}, {Romel}, {Romero}, {Romero-Shaw}, {Romie}, {Ronchini},
  {Rose}, {Rose}, {Rose}, {Rosell}, {Rosi{\'n}ska}, {Rosofsky}, {Ross},
  {Rowan}, {Rowlinson}, {Roy}, {Roy}, {Ruggi}, {Ryan}, {Sachdev}, {Sadecki},
  {Sakellariadou}, {Salafia}, {Salconi}, {Saleem}, {Samajdar}, {Sanchez},
  {Sanchez}, {Sanchez}, {Sanchis-Gual}, {Sanders}, {Santiago}, {Santos},
  {Saravanan}, {Sarin}, {Sassolas}, {Sathyaprakash}, {Sauter}, {Savage},
  {Savant}, {Sawant}, {Sayah}, {Schaetzl}, {Schale}, {Scheel}, {Scheuer},
  {Schindler-Tyka}, {Schmidt}, {Schnabel}, {Schofield}, {Sch{\"o}nbeck},
  {Schreiber}, {Schulte}, {Schutz}, {Schwarm}, {Schwartz}, {Scott}, {Scott},
  {Seglar-Arroyo}, {Seidel}, {Sellers}, {Sengupta}, {Sennett}, {Sentenac},
  {Sequino}, {Sergeev}, {Setyawati}, {Shaffer}, {Shahriar}, {Sharifi},
  {Sharma}, {Sharma}, {Shawhan}, {Shen}, {Shikauchi}, {Shink}, {Shoemaker},
  {Shoemaker}, {Shukla}, {ShyamSundar}, {Sieniawska}, {Sigg}, {Singer},
  {Singh}, {Singh}, {Singha}, {Singhal}, {Sintes}, {Sipala}, {Skliris},
  {Slagmolen}, {Slaven-Blair}, {Smetana}, {Smith}, {Smith}, {Somala}, {Son},
  {Soni}, {Sorazu}, {Sordini}, {Sorrentino}, {Sorrentino}, {Soulard},
  {Souradeep}, {Sowell}, {Spencer}, {Spera}, {Srivastava}, {Srivastava},
  {Staats}, {Stachie}, {Steer}, {Steinke}, {Steinlechner}, {Steinlechner},
  {Steinmeyer}, {Stevenson}, {Stolle-McAllister}, {Stops}, {Stover}, {Strain},
  {Stratta}, {Strunk}, {Sturani}, {Stuver}, {S{\"u}dbeck}, {Sudhagar},
  {Sudhir}, {Suh}, {Summerscales}, {Sun}, {Sun}, {Sunil}, {Sur}, {Suresh},
  {Sutton}, {Swinkels}, {Szczepa{\'n}czyk}, {Tacca}, {Tait}, {Talbot},
  {Tanasijczuk}, {Tanner}, {Tao}, {Tapia}, {Tapia San Martin}, {Tasson},
  {Taylor}, {Tenorio}, {Terkowski}, {Thirugnanasambandam}, {Thomas}, {Thomas},
  {Thomas}, {Thompson}, {Thondapu}, {Thorne}, {Thrane}, {Tiwari}, {Tiwari},
  {Tiwari}, {Toland}, {Tolley}, {Tonelli}, {Tornasi}, {Torres-Forn{\'e}},
  {Torrie}, {Tosta e Melo}, {T{\"o}yr{\"a}}, {Tran}, {Trapananti}, {Travasso},
  {Traylor}, {Tringali}, {Tripathee}, {Trovato}, {Trudeau}, {Tsai}, {Tsang},
  {Tse}, {Tso}, {Tsukada}, {Tsuna}, {Tsutsui}, {Turconi}, {Ubhi}, {Udall},
  {Ueno}, {Ugolini}, {Unnikrishnan}, {Urban}, {Usman}, {Utina}, {Vahlbruch},
  {Vajente}, {Vajpeyi}, {Valdes}, {Valentini}, {Valsan}, {van Bakel},
  {Beuzekom}, {van den Brand}, {Van Den Broeck}, {Vander-Hyde}, {van der
  Schaaf}, {van Heijningen}, {Vardaro}, {Vargas}, {Varma}, {Vass},
  {Vas{\'u}th}, {Vecchio}, {Vedovato}, {Veitch}, {Veitch}, {Venkateswara},
  {Venneberg}, {Venugopalan}, {Verkindt}, {Verma}, {Veske}, {Vetrano},
  {Vicer{\'e}}, {Viets}, {Villa-Ortega}, {Vinet}, {Vitale}, {Vo}, {Vocca},
  {Vorvick}, {Vyatchanin}, {Wade}, {Wade}, {Wade}, {Walet}, {Walker},
  {Wallace}, {Wallace}, {Walsh}, {Wang}, {Wang}, {Wang}, {Wang}, {Ward},
  {Warner}, {Was}, {Washington}, {Watchi}, {Weaver}, {Wei}, {Weinert},
  {Weinstein}, {Weiss}, {Wellmann}, {Wen}, {We{\ss}els}, {Westhouse}, {Wette},
  {Whelan}, {White}, {White}, {Whiting}, {Whittle}, {Wilken}, {Williams},
  {Williams}, {Williamson}, {Willis}, {Willke}, {Wilson}, {Wimmer}, {Winkler},
  {Wipf}, {Woan}, {Woehler}, {Wofford}, {Wong}, {Wrangel}, {Wright}, {Wu},
  {Wysocki}, {Xiao}, {Yamamoto}, {Yang}, {Yang}, {Yang}, {Yap}, {Yeeles},
  {Yoon}, {Yu}, {Yu}, {Yuen}, {Zadro{\.z}ny}, {Zanolin}, {Zelenova}, {Zendri},
  {Zevin}, {Zhang}, {Zhang}, {Zhang}, {Zhang}, {Zhao}, {Zhao}, {Zhou}, {Zhou},
  {Zhu}, {Zimmerman}, {Zucker}, {Zweizig}, {LIGO Scientific Collaboration}, \&
  {Virgo Collaboration}}]{2021ApJ...913L...7A}
{Abbott}, R., {Abbott}, T.~D., {Abraham}, S., {et~al.} 2021, \apjl, 913, L7,
  \dodoi{10.3847/2041-8213/abe949}

\bibitem[{Acernese {et~al.}(2015)}]{VIRGO:2014yos}
Acernese, F., {et~al.} 2015, Class. Quant. Grav., 32, 024001,
  \dodoi{10.1088/0264-9381/32/2/024001}

\bibitem[{Aghanim {et~al.}(2020)}]{Planck:2018vyg}
Aghanim, N., {et~al.} 2020, Astron. Astrophys., 641, A6,
  \dodoi{10.1051/0004-6361/201833910}

\bibitem[{Agrawal {et~al.}(2017)Agrawal, Makiya, Chiang, Jeong, Saito, \&
  Komatsu}]{Agrawal_2017}
Agrawal, A., Makiya, R., Chiang, C.-T., {et~al.} 2017, Journal of Cosmology and
  Astroparticle Physics, 2017, 003–003, \dodoi{10.1088/1475-7516/2017/10/003}

\bibitem[{Akutsu {et~al.}(2021)}]{KAGRA:2020tym}
Akutsu, T., {et~al.} 2021, PTEP, 2021, 05A101, \dodoi{10.1093/ptep/ptaa125}

\bibitem[{Alam {et~al.}(2021)}]{eBOSS:2020yzd}
Alam, S., {et~al.} 2021, Phys. Rev. D, 103, 083533,
  \dodoi{10.1103/PhysRevD.103.083533}

\bibitem[{Ali {et~al.}(2010)Ali, Gannouji, \& Sami}]{Ali:2010gr}
Ali, A., Gannouji, R., \& Sami, M. 2010, Phys. Rev. D, 82, 103015,
  \dodoi{10.1103/PhysRevD.82.103015}

\bibitem[{{Almeida} {et~al.}(2023){Almeida}, {Anderson},
  {Argudo-Fern{\'a}ndez}, {Badenes}, {Barger}, {Barrera-Ballesteros}, {Bender},
  {Benitez}, {Besser}, {Bizyaev}, {Blanton}, {Bochanski}, {Bovy}, {Brandt},
  {Brownstein}, {Buchner}, {Bulbul}, {Burchett}, {Cano D{\'\i}az}, {Carlberg},
  {Casey}, {Chandra}, {Cherinka}, {Chiappini}, {Coker}, {Comparat}, {Conroy},
  {Contardo}, {Cortes}, {Covey}, {Crane}, {Cunha}, {Dabbieri}, {Davidson},
  {Davis}, {De Lee}, {M{\'e}ndez Delgado}, {Demasi}, {Di Mille}, {Donor},
  {Dow}, {Dwelly}, {Eracleous}, {Eriksen}, {Fan}, {Farr}, {Frederick}, {Fries},
  {Frinchaboy}, {Gaensicke}, {Ge}, {Gonz{\'a}lez {\'A}vila}, {Grabowski},
  {Grier}, {Guiglion}, {Gupta}, {Hall}, {Hawkins}, {Hayes}, {Hermes},
  {Hern{\'a}ndez-Garc{\'\i}a}, {Hogg}, {Holtzman}, {Ibarra-Medel}, {Ji},
  {Jofre}, {Johnson}, {Jones}, {Kinemuchi}, {Kluge}, {Koekemoer}, {Kollmeier},
  {Kounkel}, {Krishnarao}, {Krumpe}, {Lacerna}, {Jakson Assuncao Lago},
  {Laporte}, {Liu}, {Liu}, {Liu}, {Lopes}, {Macktoobian}, {Malanushenko},
  {Maoz}, {Masseron}, {Masters}, {Matijevic}, {McBride}, {Medan}, {Merloni},
  {Morrison}, {Myers}, {M{\'e}sz{\'a}ros}, {Negrete}, {Nidever}, {Nitschelm},
  {Oravetz}, {Oravetz}, {Pan}, {Peng}, {Pinsonneault}, {Pogge}, {Qiu},
  {Queiroz}, {Ramirez}, {Rix}, {Fern{\'a}ndez Rosso}, {Runnoe}, {Salvato},
  {Sanchez}, {Santana}, {Saydjari}, {Sayres}, {Schlaufman}, {Schneider},
  {Schwope}, {Serna}, {Shen}, {Sobeck}, {Song}, {Souto}, {Spoo}, {Stassun},
  {Steinmetz}, {Straumit}, {Stringfellow}, {S{\'a}nchez-Gallego},
  {Taghizadeh-Popp}, {Tayar}, {Thakar}, {Tissera}, {Tkachenko}, {Hernandez
  Toledo}, {Trakhtenbrot}, {Fernandez Trincado}, {Troup}, {Trump}, {Tuttle},
  {Ulloa}, {Vazquez-Mata}, {Alfaro}, {Villanova}, {Wachter}, {Weijmans},
  {Wheeler}, {Wilson}, {Wojno}, {Wolf}, {Xue}, {Ybarra}, {Zari}, \&
  {Zasowski}}]{2023arXiv230107688A}
{Almeida}, A., {Anderson}, S.~F., {Argudo-Fern{\'a}ndez}, M., {et~al.} 2023,
  arXiv e-prints, arXiv:2301.07688, \dodoi{10.48550/arXiv.2301.07688}

\bibitem[{Bagla {et~al.}(2003)Bagla, Jassal, \& Padmanabhan}]{Bagla:2002yn}
Bagla, J.~S., Jassal, H.~K., \& Padmanabhan, T. 2003, Phys. Rev. D, 67, 063504,
  \dodoi{10.1103/PhysRevD.67.063504}

\bibitem[{Bassett \& Hlozek(2009)}]{Bassett:2009mm}
Bassett, B.~A., \& Hlozek, R. 2009.
\newblock \doarXiv{0910.5224}

\bibitem[{Baumann(2011)}]{Baumann:2009ds}
Baumann, D. 2011, in {Theoretical Advanced Study Institute in Elementary
  Particle Physics}: {Physics of the Large and the Small}, 523--686,
  \dodoi{10.1142/9789814327183_0010}

\bibitem[{Borhanian \& Sathyaprakash(2022)}]{Borhanian:2022czq}
Borhanian, S., \& Sathyaprakash, B.~S. 2022.
\newblock \doarXiv{2202.11048}

\bibitem[{Caldwell {et~al.}(1998)Caldwell, Dave, \&
  Steinhardt}]{Caldwell:1997ii}
Caldwell, R.~R., Dave, R., \& Steinhardt, P.~J. 1998, Phys. Rev. Lett., 80,
  1582, \dodoi{10.1103/PhysRevLett.80.1582}

\bibitem[{Capano {et~al.}(2020)Capano, Tews, Brown, Margalit, De, Kumar, Brown,
  Krishnan, \& Reddy}]{Capano:2019eae}
Capano, C.~D., Tews, I., Brown, S.~M., {et~al.} 2020, Nature Astron., 4, 625,
  \dodoi{10.1038/s41550-020-1014-6}

\bibitem[{Chen {et~al.}(2021)Chen, Cowperthwaite, Metzger, \&
  Berger}]{Chen:2020zoq}
Chen, H.-Y., Cowperthwaite, P.~S., Metzger, B.~D., \& Berger, E. 2021,
  Astrophys. J. Lett., 908, L4, \dodoi{10.3847/2041-8213/abdab0}

\bibitem[{Chevallier \& Polarski(2001)}]{Chevallier:2000qy}
Chevallier, M., \& Polarski, D. 2001, Int. J. Mod. Phys. D, 10, 213,
  \dodoi{10.1142/S0218271801000822}

\bibitem[{Clifton {et~al.}(2012)Clifton, Ferreira, Padilla, \&
  Skordis}]{Clifton_2012}
Clifton, T., Ferreira, P.~G., Padilla, A., \& Skordis, C. 2012, Physics
  Reports, 513, 1, \dodoi{10.1016/j.physrep.2012.01.001}

\bibitem[{Coles \& Erdogdu(2007)}]{Coles:2007be}
Coles, P., \& Erdogdu, P. 2007, JCAP, 10, 007,
  \dodoi{10.1088/1475-7516/2007/10/007}

\bibitem[{Dalal {et~al.}(2006)Dalal, Holz, Hughes, \& Jain}]{Dalal:2006qt}
Dalal, N., Holz, D.~E., Hughes, S.~A., \& Jain, B. 2006, Phys. Rev. D, 74,
  063006, \dodoi{10.1103/PhysRevD.74.063006}

\bibitem[{Eisenstein \& Hu(1998)}]{Eisenstein:1997ik}
Eisenstein, D.~J., \& Hu, W. 1998, Astrophys. J., 496, 605,
  \dodoi{10.1086/305424}

\bibitem[{Eisenstein {et~al.}(2005)}]{SDSS:2005xqv}
Eisenstein, D.~J., {et~al.} 2005, Astrophys. J., 633, 560,
  \dodoi{10.1086/466512}

\bibitem[{{Evans} {et~al.}(2021){Evans}, {Adhikari}, {Afle}, {Ballmer},
  {Biscoveanu}, {Borhanian}, {Brown}, {Chen}, {Eisenstein}, {Gruson}, {Gupta},
  {Hall}, {Huxford}, {Kamai}, {Kashyap}, {Kissel}, {Kuns}, {Landry}, {Lenon},
  {Lovelace}, {McCuller}, {Ng}, {Nitz}, {Read}, {Sathyaprakash}, {Shoemaker},
  {Slagmolen}, {Smith}, {Srivastava}, {Sun}, {Vitale}, \&
  {Weiss}}]{2021arXiv210909882E}
{Evans}, M., {Adhikari}, R.~X., {Afle}, C., {et~al.} 2021, arXiv e-prints,
  arXiv:2109.09882.
\newblock \doarXiv{2109.09882}

\bibitem[{Fairhurst(2014)}]{Fairhurst:2012tf}
Fairhurst, S. 2014, J. Phys. Conf. Ser., 484, 012007,
  \dodoi{10.1088/1742-6596/484/1/012007}

\bibitem[{Gannouji \& Sami(2010)}]{Gannouji:2010au}
Gannouji, R., \& Sami, M. 2010, Phys. Rev. D, 82, 024011,
  \dodoi{10.1103/PhysRevD.82.024011}

\bibitem[{Giostri {et~al.}(2012)Giostri, dos Santos, Waga, Reis, Calv\~ao, \&
  Lago}]{Giostri:2012ek}
Giostri, R., dos Santos, M.~V., Waga, I., {et~al.} 2012, JCAP, 03, 027,
  \dodoi{10.1088/1475-7516/2012/03/027}

\bibitem[{Guth(1981)}]{Guth:1980zm}
Guth, A.~H. 1981, Phys. Rev. D, 23, 347, \dodoi{10.1103/PhysRevD.23.347}

\bibitem[{Holz \& Hughes(2005)}]{Holz:2005df}
Holz, D.~E., \& Hughes, S.~A. 2005, Astrophys. J., 629, 15,
  \dodoi{10.1086/431341}

\bibitem[{Hu \& Dodelson(2002)}]{Hu:2001bc}
Hu, W., \& Dodelson, S. 2002, Ann. Rev. Astron. Astrophys., 40, 171,
  \dodoi{10.1146/annurev.astro.40.060401.093926}

\bibitem[{Ivezi\'c {et~al.}(2019)}]{LSST:2008ijt}
Ivezi\'c, v., {et~al.} 2019, Astrophys. J., 873, 111,
  \dodoi{10.3847/1538-4357/ab042c}

\bibitem[{Kumar {et~al.}(2022)Kumar, Vijaykumar, \& Nitz}]{Kumar:2021aog}
Kumar, S., Vijaykumar, A., \& Nitz, A.~H. 2022, Astrophys. J., 930, 113,
  \dodoi{10.3847/1538-4357/ac5e34}

\bibitem[{Landy \& Szalay(1993)}]{Landy:1993yu}
Landy, S.~D., \& Szalay, A.~S. 1993, Astrophys. J., 412, 64,
  \dodoi{10.1086/172900}

\bibitem[{Laureijs {et~al.}(2011)Laureijs, Amiaux, Arduini, Auguères,
  Brinchmann, Cole, Cropper, Dabin, Duvet, Ealet, Garilli, Gondoin, Guzzo,
  Hoar, Hoekstra, Holmes, Kitching, Maciaszek, Mellier, Pasian, Percival,
  Rhodes, Criado, Sauvage, Scaramella, Valenziano, Warren, Bender, Castander,
  Cimatti, Fèvre, Kurki-Suonio, Levi, Lilje, Meylan, Nichol, Pedersen, Popa,
  Lopez, Rix, Rottgering, Zeilinger, Grupp, Hudelot, Massey, Meneghetti,
  Miller, Paltani, Paulin-Henriksson, Pires, Saxton, Schrabback, Seidel, Walsh,
  Aghanim, Amendola, Bartlett, Baccigalupi, Beaulieu, Benabed, Cuby, Elbaz,
  Fosalba, Gavazzi, Helmi, Hook, Irwin, Kneib, Kunz, Mannucci, Moscardini, Tao,
  Teyssier, Weller, Zamorani, Osorio, Boulade, Foumond, Di~Giorgio, Guttridge,
  James, Kemp, Martignac, Spencer, Walton, Blümchen, Bonoli, Bortoletto,
  Cerna, Corcione, Fabron, Jahnke, Ligori, Madrid, Martin, Morgante, Pamplona,
  Prieto, Riva, Toledo, Trifoglio, Zerbi, Abdalla, Douspis, Grenet, Borgani,
  Bouwens, Courbin, Delouis, Dubath, Fontana, Frailis, Grazian, Koppenhöfer,
  Mansutti, Melchior, Mignoli, Mohr, Neissner, Noddle, Poncet, Scodeggio,
  Serrano, Shane, Starck, Surace, Taylor, Verdoes-Kleijn, Vuerli, Williams,
  Zacchei, Altieri, Sanz, Kohley, Oosterbroek, Astier, Bacon, Bardelli, Baugh,
  Bellagamba, Benoist, Bianchi, Biviano, Branchini, Carbone, Cardone, Clements,
  Colombi, Conselice, Cresci, Deacon, Dunlop, Fedeli, Fontanot, Franzetti,
  Giocoli, Garcia-Bellido, Gow, Heavens, Hewett, Heymans, Holland, Huang,
  Ilbert, Joachimi, Jennins, Kerins, Kiessling, Kirk, Kotak, Krause, Lahav, van
  Leeuwen, Lesgourgues, Lombardi, Magliocchetti, Maguire, Majerotto, Maoli,
  Marulli, Maurogordato, McCracken, McLure, Melchiorri, Merson, Moresco,
  Nonino, Norberg, Peacock, Pello, Penny, Pettorino, Di~Porto, Pozzetti,
  Quercellini, Radovich, Rassat, Roche, Ronayette, Rossetti, Sartoris,
  Schneider, Semboloni, Serjeant, Simpson, Skordis, Smadja, Smartt, Spano,
  Spiro, Sullivan, Tilquin, Trotta, Verde, Wang, Williger, Zhao, Zoubian, \&
  Zucca}]{euclidmission}
Laureijs, R., Amiaux, J., Arduini, S., {et~al.} 2011, Euclid Definition Study
  Report,  arXiv, \dodoi{10.48550/ARXIV.1110.3193}

\bibitem[{Linde(1982)}]{Linde:1981mu}
Linde, A.~D. 1982, Phys. Lett. B, 108, 389,
  \dodoi{10.1016/0370-2693(82)91219-9}

\bibitem[{Mastrogiovanni {et~al.}(2021)Mastrogiovanni, Leyde, Karathanasis,
  Chassande-Mottin, Steer, Gair, Ghosh, Gray, Mukherjee, \&
  Rinaldi}]{Mastrogiovanni:2021wsd}
Mastrogiovanni, S., Leyde, K., Karathanasis, C., {et~al.} 2021, Phys. Rev. D,
  104, 062009, \dodoi{10.1103/PhysRevD.104.062009}

\bibitem[{Mills {et~al.}(2018)Mills, Tiwari, \& Fairhurst}]{Mills:2017urp}
Mills, C., Tiwari, V., \& Fairhurst, S. 2018, Phys. Rev., D97, 104064,
  \dodoi{10.1103/PhysRevD.97.104064}

\bibitem[{Mukherjee {et~al.}(2021)Mukherjee, Wandelt, Nissanke, \&
  Silvestri}]{Mukherjee:2020hyn}
Mukherjee, S., Wandelt, B.~D., Nissanke, S.~M., \& Silvestri, A. 2021, Phys.
  Rev. D, 103, 043520, \dodoi{10.1103/PhysRevD.103.043520}

\bibitem[{Nicolis {et~al.}(2009)Nicolis, Rattazzi, \&
  Trincherini}]{Nicolis:2008in}
Nicolis, A., Rattazzi, R., \& Trincherini, E. 2009, Phys. Rev. D, 79, 064036,
  \dodoi{10.1103/PhysRevD.79.064036}

\bibitem[{Nissanke {et~al.}(2013)Nissanke, Holz, Dalal, Hughes, Sievers, \&
  Hirata}]{Nissanke:2013fka}
Nissanke, S., Holz, D.~E., Dalal, N., {et~al.} 2013.
\newblock \doarXiv{1307.2638}

\bibitem[{Nitz \& Dal~Canton(2021)}]{Nitz_2021}
Nitz, A.~H., \& Dal~Canton, T. 2021, The Astrophysical Journal Letters, 917,
  L27, \dodoi{10.3847/2041-8213/ac1a75}

\bibitem[{Nitz {et~al.}(2021)Nitz, Kumar, Wang, Kastha, Wu, Schäfer,
  Dhurkunde, \& Capano}]{nitz20214ogc}
Nitz, A.~H., Kumar, S., Wang, Y.-F., {et~al.} 2021, 4-OGC: Catalog of
  gravitational waves from compact-binary mergers.
\newblock \doarXiv{2112.06878}

\bibitem[{{Peebles}(1980)}]{1980lssu.book.....P}
{Peebles}, P.~J.~E. 1980, {The large-scale structure of the universe}

\bibitem[{Peebles \& Ratra(1988)}]{Peebles:1987ek}
Peebles, P. J.~E., \& Ratra, B. 1988, Astrophys. J. Lett., 325, L17,
  \dodoi{10.1086/185100}

\bibitem[{{Petrov} {et~al.}(2022){Petrov}, {Singer}, {Coughlin}, {Kumar},
  {Almualla}, {Anand}, {Bulla}, {Dietrich}, {Foucart}, \&
  {Guessoum}}]{2022ApJ...924...54P}
{Petrov}, P., {Singer}, L.~P., {Coughlin}, M.~W., {et~al.} 2022, \apj, 924, 54,
  \dodoi{10.3847/1538-4357/ac366d}

\bibitem[{Punturo {et~al.}(2010)Punturo, Abernathy, Acernese, Allen, Andersson,
  Arun, Barone, Barr, Barsuglia, Beker, Beveridge, Birindelli, Bose, Bosi,
  Braccini, Bradaschia, Bulik, Calloni, Cella, Mottin, Chelkowski, Chincarini,
  Clark, Coccia, Colacino, Colas, Cumming, Cunningham, Cuoco, Danilishin,
  Danzmann, Luca, Salvo, Dent, Rosa, Fiore, Virgilio, Doets, Fafone, Falferi,
  Flaminio, Franc, Frasconi, Freise, Fulda, Gair, Gemme, Gennai, Giazotto,
  Glampedakis, Granata, Grote, Guidi, Hammond, Hannam, Harms, Heinert, Hendry,
  Heng, Hennes, Hild, Hough, Husa, Huttner, Jones, Khalili, Kokeyama, Kokkotas,
  Krishnan, Lorenzini, Lück, Majorana, Mandel, Mandic, Martin, Michel,
  Minenkov, Morgado, Mosca, Mours, Müller–Ebhardt, Murray, Nawrodt, Nelson,
  Oshaughnessy, Ott, Palomba, Paoli, Parguez, Pasqualetti, Passaquieti,
  Passuello, Pinard, Poggiani, Popolizio, Prato, Puppo, Rabeling, Rapagnani,
  Read, Regimbau, Rehbein, Reid, Rezzolla, Ricci, Richard, Rocchi, Rowan,
  Rüdiger, Sassolas, Sathyaprakash, Schnabel, Schwarz, Seidel, Sintes, Somiya,
  Speirits, Strain, Strigin, Sutton, Tarabrin, Thüring, van~den Brand, van
  Leewen, van Veggel, van~den Broeck, Vecchio, Veitch, Vetrano, Vicere,
  Vyatchanin, Willke, Woan, Wolfango, \& Yamamoto}]{Punturo_2010}
Punturo, M., Abernathy, M., Acernese, F., {et~al.} 2010, Classical and Quantum
  Gravity, 27, 194002, \dodoi{10.1088/0264-9381/27/19/194002}

\bibitem[{Ratra \& Peebles(1988)}]{Ratra:1987rm}
Ratra, B., \& Peebles, P. J.~E. 1988, Phys. Rev. D, 37, 3406,
  \dodoi{10.1103/PhysRevD.37.3406}

\bibitem[{Reitze {et~al.}(2019)}]{Reitze:2019iox}
Reitze, D., {et~al.} 2019, Bull. Am. Astron. Soc., 51, 035.
\newblock \doarXiv{1907.04833}

\bibitem[{Riess {et~al.}(2019)Riess, Casertano, Yuan, Macri, \&
  Scolnic}]{Riess:2019cxk}
Riess, A.~G., Casertano, S., Yuan, W., Macri, L.~M., \& Scolnic, D. 2019,
  Astrophys. J., 876, 85, \dodoi{10.3847/1538-4357/ab1422}

\bibitem[{Riess \& et~al.(1998)}]{riess1998observational}
Riess, A.~G., \& et~al. 1998, The Astronomical Journal, 116, 1009,
  \dodoi{10.1086/300499}

\bibitem[{Riess {et~al.}(2021)}]{Riess:2021jrx}
Riess, A.~G., {et~al.} 2021.
\newblock \doarXiv{2112.04510}

\bibitem[{Saleem {et~al.}(2022)}]{Saleem:2021iwi}
Saleem, M., {et~al.} 2022, Class. Quant. Grav., 39, 025004,
  \dodoi{10.1088/1361-6382/ac3b99}

\bibitem[{Sathyaprakash {et~al.}(2012)}]{Sathyaprakash:2012jk}
Sathyaprakash, B., {et~al.} 2012, Class. Quant. Grav., 29, 124013,
  \dodoi{10.1088/010264-9381/29/12/124013}

\bibitem[{Speagle(2020)}]{speagle:2019}
Speagle, J.~S. 2020, Monthly Notices of the Royal Astronomical Society, 493,
  3132, \dodoi{10.1093/mnras/staa278}

\bibitem[{Turner \& White(1997)}]{Turner:1997npq}
Turner, M.~S., \& White, M.~J. 1997, Phys. Rev. D, 56, R4439,
  \dodoi{10.1103/PhysRevD.56.R4439}

\bibitem[{Vijaykumar {et~al.}(2020)Vijaykumar, Saketh, Kumar, Ajith, \&
  Choudhury}]{Vijaykumar:2020pzn}
Vijaykumar, A., Saketh, M. V.~S., Kumar, S., Ajith, P., \& Choudhury, T.~R.
  2020.
\newblock \doarXiv{2005.01111}

\bibitem[{{Weinberg}(2013)}]{weinberg2013observational}
{Weinberg}, D.~H. 2013, Physics Reports, 530, 87,
  \dodoi{10.1016/j.physrep.2013.05.001}

\bibitem[{Weinberg {et~al.}(2013)Weinberg, Mortonson, Eisenstein, Hirata,
  Riess, \& Rozo}]{Weinberg:2013agg}
Weinberg, D.~H., Mortonson, M.~J., Eisenstein, D.~J., {et~al.} 2013, Phys.
  Rept., 530, 87, \dodoi{10.1016/j.physrep.2013.05.001}

\bibitem[{Wetterich(1988)}]{Wetterich:1987fm}
Wetterich, C. 1988, Nucl. Phys. B, 302, 668,
  \dodoi{10.1016/0550-3213(88)90193-9}

\bibitem[{Zlatev {et~al.}(1999)Zlatev, Wang, \& Steinhardt}]{Zlatev:1998tr}
Zlatev, I., Wang, L.-M., \& Steinhardt, P.~J. 1999, Phys. Rev. Lett., 82, 896,
  \dodoi{10.1103/PhysRevLett.82.896}

\end{thebibliography}

\end{CJK*}
\end{document}